\documentclass[aps,prl,twocolumn,superscriptaddress,floatfix]{revtex4-1}

\usepackage{physics}
\usepackage{graphicx}
\usepackage{booktabs}
\usepackage{bm, bbm, amssymb, siunitx}
\renewcommand{\figurename}{{\bf Fig.}}

\newcommand{\smb}{\textrm{SmB}_6}
\newcounter{para}
\newcommand\mypara{ \par\refstepcounter{para}\noindent \textbf{\thepara}\indent}

\usepackage{hyperref}
\hypersetup{
    colorlinks=true,
    linkcolor=blue,     
    urlcolor=blue,
    citecolor=blue
}
\setcitestyle{round}

\begin{document}

\title{Imaging emergent heavy Dirac fermions of a topological Kondo insulator}

\author{Harris Pirie}
\author{Yu Liu}
\author{A. Soumyanarayanan}
\author{Pengcheng Chen}
\author{Yang He}
\author{M.M. Yee}
\affiliation{Department of Physics, Harvard University, Cambridge, MA, 02138, USA}
\author{P.F.S. Rosa}
\author{J.D. Thompson}
\affiliation{Los Alamos National Laboratory, Los Alamos, NM 87545, USA}
\author{Dae-Jeong Kim}
\author{Z. Fisk}
\affiliation{Department of Physics and Astronomy, University of California at Irvine, Irvine, CA 92697, USA}
\author{Xiangfeng Wang}
\author{J. Paglione}
\affiliation{Center for Nanophysics and Advanced Materials, Department of Physics, University of Maryland, College Park, Maryland 20742, USA}
\author{Dirk K. Morr}
\affiliation{Department of Physics, University of Illinois at Chicago, Chicago, IL, 60607, USA}
\author{M. H. Hamidian}
\affiliation{Department of Physics, Harvard University, Cambridge, MA, 02138, USA}
\author{Jennifer E. Hoffman}
\affiliation{Department of Physics, Harvard University, Cambridge, MA, 02138, USA}

\date{\today}

\begin{abstract}
Kondo insulators are primary candidates in the search for strongly correlated topological quantum phases, which may host topological order, fractionalization, and non-Abelian statistics. Within some Kondo insulators, the hybridization gap is predicted to protect a nontrivial topological invariant and to harbor emergent heavy Dirac fermion surface modes. We use high-energy-resolution spectroscopic imaging in real and momentum space on the Kondo insulator, $\smb$.  On cooling through $T^*_{\Delta}\approx$ 35 K we observe the opening of an insulating gap that expands to $\Delta\approx$ 10 meV at 2 K.  Within the gap, we image the formation of linearly dispersing surface states with effective masses reaching $m^* = (410\pm20)m_e$. We thus demonstrate existence of a strongly correlated topological Kondo insulator phase hosting the heaviest known Dirac fermions.
\end{abstract}

 
\maketitle


\begin{figure}[h!]
	\includegraphics[width=0.5\textwidth]{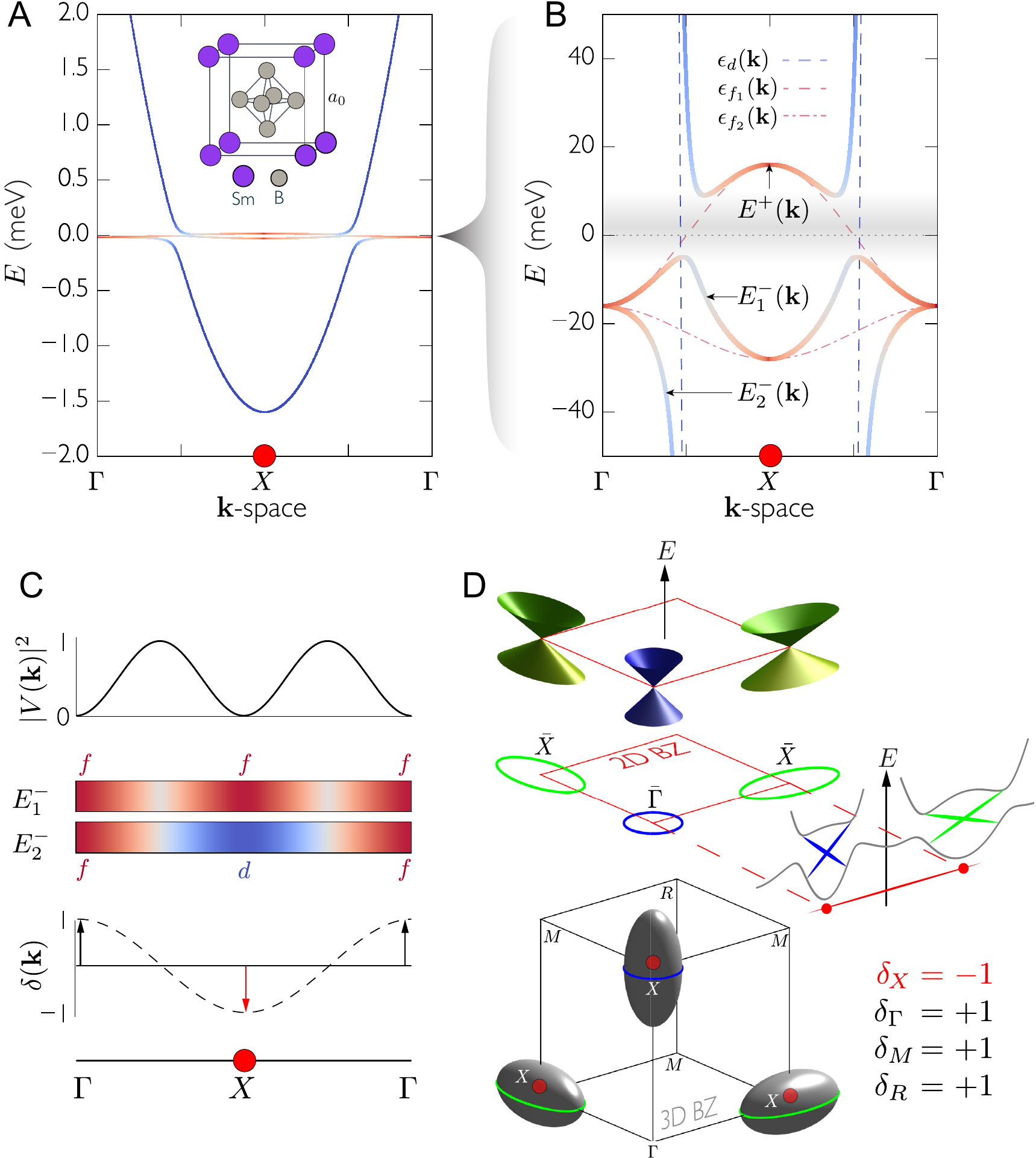}
	\caption{\label{F1}{\bf Anticipated topological Kondo insulator electronic structure of $\smb$.}
            ({\bf A}) The Kondo insulator (KI) electronic structure consists of an itinerant $d$-character band (blue), centered around the $X$ point, which hybridizes with localized $f$-character (red) states. Both $f$ and $d$ states are contributed by the Sm atoms, which form a cubic unit cell (inset).
            ({\bf B}) Narrow energy window of the same electronic structure in (A). As the temperature is lowered one $d$-band, $\epsilon_d$, and two closely spaced crystal-field-split $f$-states, $\epsilon_{f_1}, \epsilon_{f_2}$, hybridize to form three separate bands, $E^+,E_1^{-}, E_2^{-}$, with a gap of several meV.   
            ({\bf C}) The topological invariant for the KI electronic structure is calculated from a product of parity eigenvalues, which are opposite for $d$ and $f$ states. Top: Nodes in the hybridization parameter, $|V(\mathbf{k})|^2$, at the $X$ and $\Gamma$ points lead to pure $f$ or $d$ character of the hybridized bands at those locations. Middle:  The filled bands, $E_1^{-}, E_2^{-}$ have full $f$ character at $\Gamma$ and evolve to either $d$ or $f$ at $X$. Bottom: Thus, total parity, $\delta$, is reversed only at the $X$ points (marked red circle).
            ({\bf D}) The cubic topological Kondo insulator (TKI) electronic structure has parity inversion at three $X$ points (red balls) in the 3D  Brillouin zone giving a $Z_2$ topological index $\nu =\delta_{\Gamma} \delta_{R} (\delta_{X} \delta_{M})^3 = -1$ that encodes the strong topological state.  When projected onto the 2D Brillouin zone, the inversion manifests at the $\bar{\Gamma}$ and two $\bar{X}$ points (red circles) and consequently sets the locations of the predicted 2D Dirac states. These Dirac surface states acquire different velocities because the three symmetry-equivalent constant energy contours have inequivalent projects onto the 2D surface Brillouin zone (green ellipses and blue circle).}
\end{figure}

\begin{figure*}
	\includegraphics[width=0.915\textwidth]{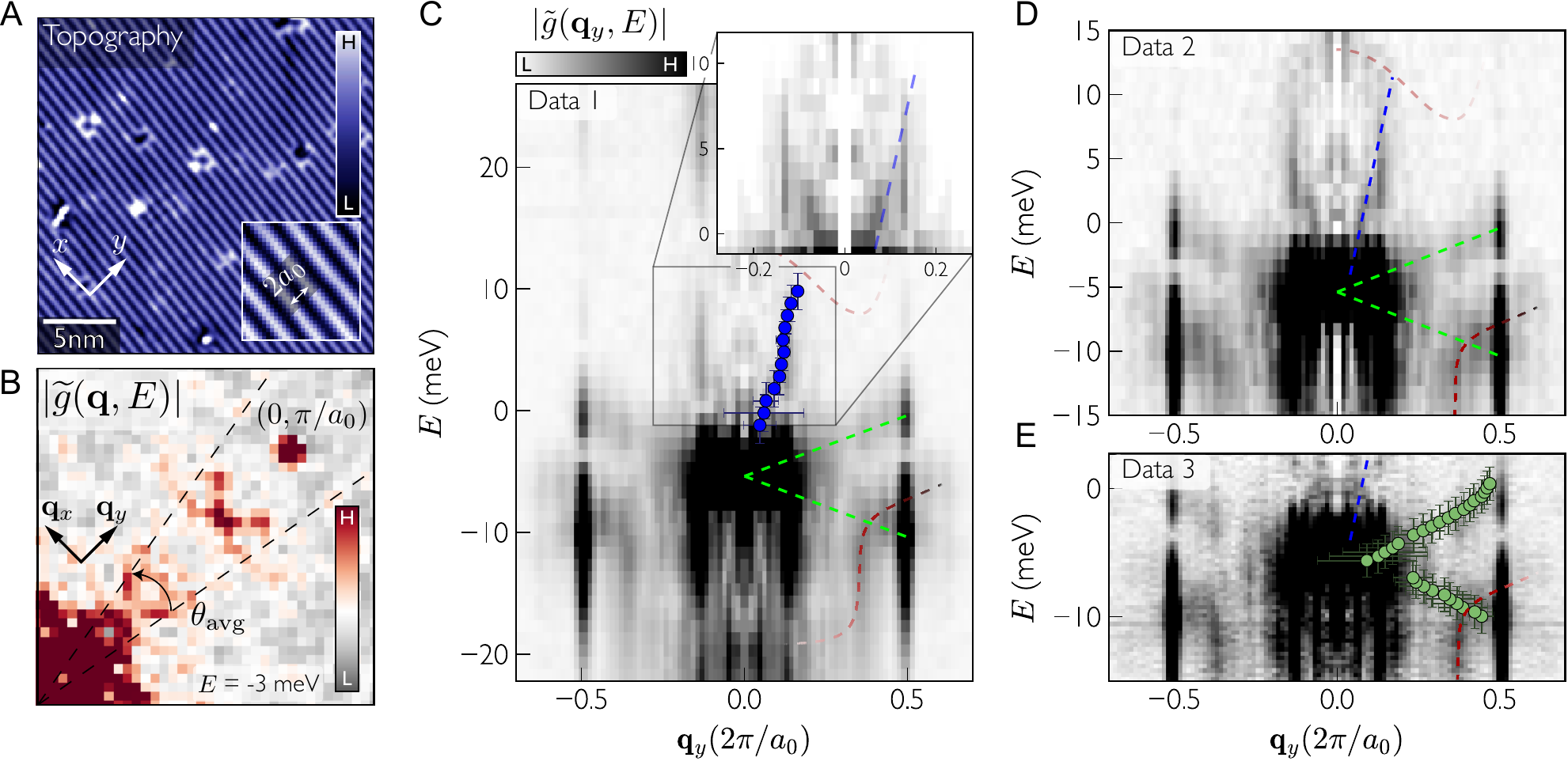}
	\caption{\label{F2}{\bf Raw quasiparticle interference (QPI) from three different sample areas.}
            ({\bf A})  A $(2\times 1)$ surface reconstruction on the half-Sm termination of $\smb$ doubles the unit cell in the $y$ direction, creating the rows of atoms shown in the topographic image (sample bias $V_\mathrm{s}= -50$ meV, current set point $I_\mathrm{s} =  100$ pA).
		    ({\bf B}) Fourier transform (FT) magnitude of spatially resolved differential conductance maps, $|\tilde{g}(\mathbf{q},E)|$ for $E = -2$ meV, acquired in a typical field of view as shown in (A). Scattering of momentum eigenstates from defects generates quasiparticle interference (QPI) manifesting as peaks in $\mathbf{q}$-space whose trajectory is determined by $\mathbf{k}$-space electronic structure.  The largest signal appears along the $\mathbf{q}_y$ direction. The image has been two-fold symmetrized to increase signal-to-noise.
            ({\bf C-E}) $\theta$-averaged linecuts along $\mathbf{q}_y$ of $|\tilde{g}(\mathbf{q},E)|$ on three different sample areas give consistent results. In each case, dispersing QPI signals are marked by dashed guides: blue and green lines track surface states, while the red guides track KI states (section III of \cite{Supp}).  The surface states are quantified by fitting each row to a sum of Gaussians that reflect contributions from the Bragg peak, low-$\mathbf{q}$ disorder and dispersing QPI (circles). A box-windowed FT (zoomed inset in (C)) enhances the low-$\mathbf{q}$ signal (blue) compared to the Hanning-windowed FT that reduces spread of the high-$\mathbf{q}$ states in the main panel. The $\mathbf{q}$-axis error bars are estimates based on the covariance matrix of the Gaussian fits, whereas the $E$-axis error bars show the energy resolution of the STM. 
            }
\end{figure*}


\begin{figure*}
	\includegraphics[width=0.85\textwidth]{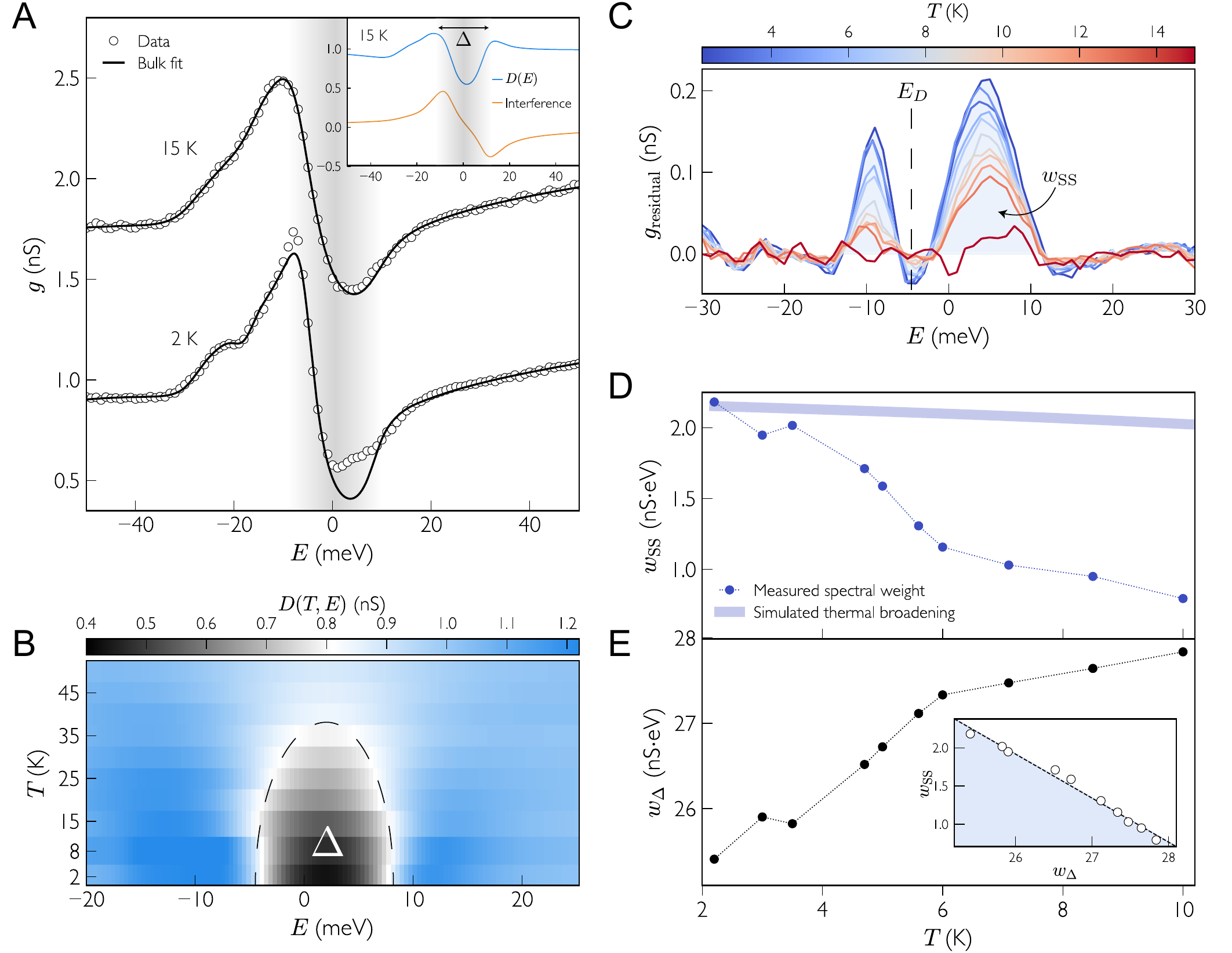}
	\caption{\label{F4}{\bf Concomitant evolution of topological Dirac states and         KI gap.}
            ({\bf A})  Quantum mechanical interference of electrons tunneling from the tip to either $d$ or $f$ states in the $\smb$ Kondo lattice renders $g(E)$ data (open circles) inequivalent to the density of states. The data must be fit to a co-tunneling model  representing bulk contributions (equation (\ref{eq:didv}), black line) from which the KI density of states can be extracted. The fit at 15 K captures nearly all features of the spectrum: the inset shows KI state contributions, $D(E)$ (blue), when the interference terms (orange) are removed. At 2 K, the fit deviates for energies within the KI gap (grey region) signaling that the bulk KI structure in the co-tunneling fit no longer fully captures the data.
            ({\bf B}) The temperature evolution of KI states in $\smb$ from fitting $g(E, T)$ to a co-tunneling model (see blue curve in (A)). The KI gap $\Delta$ is observable below $T^*_{\Delta}\approx 35$ K and reaches $\approx 10$ meV at low temperatures.  
            ({\bf C}) Subtraction of co-tunneling model from $g(E)$ for a sequence of temperatures. The intensifying V-shaped residuals are characteristic of a linear band with a nodal energy at $-5$ meV, in agreement with our independent raw QPI measurements showing Dirac states (see Figs.\ \ref{F2}C-E). 
            ({\bf D}) As temperature increases, the integrated spectral weight under the residuals in (C), $w_{\mathrm{SS}}$ blue dots, decays faster than expected from thermal broadening (thick blue line, see section VI of \cite{Supp}).
            ({\bf E}) The integrated spectral weight within $\Delta$, $w_{\Delta}$ black dots, shows a sharpening of the KI gap at lower temperatures. The formation of the Dirac states observed in (D) mirrors the sharpening of the KI gap at low temperatures, demonstrating the direct relationship between the evolving host insulator and its topologically emergent states (see inset).  
            }
\end{figure*}

\mypara
Correlated topological matter is a frontier in the search for exotic quantum phases. Heavy fermion systems were recently predicted to host a novel \emph{topological Kondo insulators} (TKI) phase \cite{Dzero2010}. Kondo insulators (KI) \cite{Dzero2016} are formed when strong interactions within a periodic array of localized moments, usually $f$-electrons, lead to reorganization of the low temperature electronic structure.  The process opens an insulating gap $\Delta$, driven by hybridization between renormalized, low-lying localized states and itinerant conduction electrons (Figs.\ \ref{F1}A-B). The KI $\smb$, in which localized $f$ and itinerant $d$ states are contributed by Sm atoms (Fig.\ \ref{F1}A inset), exhibits such a characteristic metal-insulator crossover at $T_{MI}\approx$ 30-50 K \cite{Allen1979,Kim2014}. Mysteriously, the resistance plateaus below 5 K \cite{Allen1979,Kim2014}, signaling the onset of a new conduction channel whose origin has been intensely debated for decades \cite{Dzero2016}. One resolution to the long-standing problem posits that the $\smb$ ground state is a TKI \cite{Dzero2010,Alexandrov2013,Takimoto2011,Lu2013}, whose low temperature conductivity originates from topologically emergent Dirac surface states within the narrow energy window of the KI gap.   Large $f$-electron contributions to these states are predicted to yield the heaviest Dirac states of any known material. 

\mypara
Although recently discovered topological insulators have reasserted the importance of topology in condensed matter systems, the topological invariant of their bulk band structure is fully understood within a \emph{non-interacting} fermion description, and their surface states are trivial metals with ordinary fermionic excitations \cite{Ando2013}.  In contrast, the many-body ground states of \emph{interacting} or strongly correlated systems are expected to generate long-range entanglement and ground state degeneracy, leading to topological \emph{order} \cite{Wen1990}.  Furthermore, predictions for surface states of strongly correlated topological materials include interacting Dirac liquids with spontaneously generated anomalous quantum Hall states \cite{Efimkin2014}, topological order \cite{Wang2013,Chen2014,Thomson2016}, fractionalization \cite{Thomson2016}, and non-Abelian exchange statistics \cite{Wang2013,Chen2014}.  The prodigious density of the Dirac states expected near the Fermi level magnifies their susceptibility to some of these anticipated novel orders, and their potential utility. 

\mypara
Electronic structure calculations \cite{Takimoto2011,Alexandrov2013,Lu2013} predict $\smb$ to be a TKI. Figs.\ \ref{F1}A,B show the expected bulk band structure of the associated Kondo lattice model with two crystal-field-split $f$ states and a band of $d$ electrons.  Quantum mechanical mixing between opposite parity $f$ and $d$ states vanishes at the high symmetry points $\mathbf{k}^* =\Gamma,X,M,R$ where parity is a good quantum number, thus forcing nodes in the hybridization matrix element, $\bra{f}V(\mathbf{k}^*)\ket{d}=0$ (Fig.\ \ref{F1}C). Consequently, the parity $\delta(\mathbf{k})$ of the fully hybridized filled states is inverted at three symmetry-equivalent $X$-points in the 3D cubic Brillouin zone.  The resulting topological invariant \cite{Dzero2010,Alexandrov2013}, captured by the index $\nu =\delta_{\Gamma}\delta_{R}(\delta_{X}\delta_{M})^3 = -1$, where $\delta_{\Gamma,X,M,R} = \pm 1$ is the parity of filled bands at high symmetry points (red circles Fig.\ \ref{F1}D), predicts a non-trivial topological phase.  On the $\smb$ Kondo lattice, topologically emergent surface states are predicted to i) lie predominantly within the energy interval of the bulk gap generated by correlations, ii) have a Dirac spectrum, iii) be centered at the $\bar{\Gamma}$ and two $\bar{X}$-points of the 2D surface Brillouin zone, and iv) have distinct band velocities.  The last prediction follows intuitively from observing that 3D constant-energy manifolds at the $X$-points with different relative orientations create inequivalent 2D projections (Fig.\ \ref{F1}D).  Consequently, surface states connecting 2D projected bulk hybridized bands are expected to have distinct velocities at the $\bar{\Gamma}$ and $\bar{X}$-points (Fig.\ \ref{F1}D).


\mypara
Experimentally, temperature-dependent point-contact spectroscopy \cite{Zhang2013} has shown the $\smb$ bulk electronic structure to be consistent with a KI, while surface conduction channels have been demonstrated by geometry-dependent transport measurements at low temperatures \cite{Kim2014,Syers2015}.  Magnetothermoelectric studies \cite{Luo2015} imply that the $(110)$ surface of $\smb$ may harbor heavy metallic surface states, though the topological nature of such states is not clear.  On the other hand, torque magnetometry experiments on the $(110)$ surface measured surface states whose velocities are more than two orders of magnitude larger \cite{Li2014} than expected for $\smb$ surface states \cite{Lu2013}. In contrast, a separate quantum oscillation measurement attributed these states to the bulk \cite{Tan2015}.  Direct energy and momentum resolved electronic structure mapping by angle resolved photoemission spectroscopy (ARPES) \cite{Xu2014,Jiang2013,Neupane2013} appeared to show linearly dispersing surface bands with velocities at least an order of magnitude larger than those expected for the Dirac fermions of $\smb$ and with an inferred Dirac point buried far below the bulk gap \cite{Lu2013} (table II in \cite{Supp}).  Because much of the active physics in the ground state resides within a small energy window set by the KI gap, $\Delta \approx$ 8-10 meV, as revealed by transport experiments \cite{Gorshunov1999,Syers2015}, ARPES mapping of the electronic structure is limited in detecting narrow bandwidth states of a incipient TKI \cite{Frantzeskakis2013}.  Furthermore, ARPES measurements typically average over different (polar) surface terminations with relative intensity and chemical potential shifts in their surface states, rendering interpretations difficult \cite{Hlawenka2018}.  Collectively, these suggestive but controversial experiments have renewed the urgency to discover topological states arising from strong electronic interactions.  Ultimately, observation of strongly correlated topological states in $\smb$ requires measurements on a uniform and ordered surface termination, access to filled and empty states at low temperatures, and meV energy resolution in momentum space to disentangle the shallow dispersions of a bulk KI band structure (see Fig.\ \ref{F1}B) and surface heavy Dirac fermions (see Fig.\ \ref{F1}D). 

\mypara
To search for a strongly correlated topological phase in $\smb$ and the topologically emergent heavy Dirac fermions, we used spectroscopic scanning tunneling microscopy (STM) to image the temperature-dependent electronic structure in both $\mathbf{r}$ and $\mathbf{k}$-space. We studied the (001) surface of $\smb$, prepared by cleaving single crystals in cryogenic ultra-high vacuum and directly inserting into the STM at 4 K. We focus on regions where exactly half of the Sm atoms remain on the cleaved surface, resulting in an ordered $(2\times 1)$ reconstruction (Fig.\ \ref{F2}A) \cite{Ruan2014,Roßler2014} that is non-polar and thus eliminates the possibility of polarity-driven surface states \cite{Zhu2013}. We measured energy-resolved differential conductance $dI/dV(\mathbf{r},E = eV) \equiv g(\mathbf{r},E)$, where $I$ is the tunneling current and $V$ is the bias applied between the sample and STM tip. We used 1.5 meV energy resolution in six separate fields of view, each $\sim30$ nm in size. In each case, the periodic modulations in $g(\mathbf{r},E)$ around the defect sites are identified as quasiparticle interference (QPI) patterns, generated by elastic scattering of $\mathbf{k}$-space eigenstates.  The energy-resolved interference modulation wavevector, $\mathbf{q}(E) = \mathbf{k}_f(E)-\mathbf{k}_i(E)$ encodes the crystal momentum transfer between initial $(\mathbf{k}_i)$ and final $(\mathbf{k}_i)$ states \cite{Schmidt2010,Aynajian2012,Allan2013}.  The modulations manifest as peaks in the Fourier transform (Fig.\ \ref{F2}B) whose energy-dependent trajectory (representative raw data in Figs.\ \ref{F2}C-E) can be used to infer \textbf{k}-space electronic structure (see section III of \cite{Supp} for additional raw data and analysis).


\mypara
Dispersing QPI trajectories corresponding to distinct components of the $\smb$ electronic structure are observable in the Fourier transform of $g(\mathbf{r},E)$, with highest signal-to-noise along $\mathbf{q}_y$ (Figs.\ \ref{F2}C-E) primarily due to the anisotropy of the scattering form factor \cite{Allan2013a}.  Two sets of dispersions, one very shallow (green guides) and one steep (blue guides), are observed at energies within the Kondo insulator gap, $\Delta$.  The remaining set of dispersions, observed at higher $|E|$ (dashed red guides), can be mapped to the known low-energy Kondo insulator states of $\smb$ (see section III of \cite{Supp}). The dispersions are reproducible in six different raw datasets, on three distinct samples, from two different growers, with distinct STM acquisition parameters, three of which are shown in Figs.\ \ref{F2}C-E. We quantify these in-gap states by fitting each energy of $|\tilde{g}(\mathbf{q}_y, E)|$ with a Gaussian or sum of Gaussians to distinguish the contributions from dispersing QPI, static low-$\mathbf{q}$ disorder and Bragg peaks (details in  section III of \cite{Supp}). The extracted dispersions of scattering vectors in the QPI, overlaid on Figs.\ \ref{F2}C and E, show two sets of linear traces that would be expected from scattering within distinct Dirac cones  (section III and Figs.\ S\ref{FGQPI}-\ref{FXQPI} in \cite{Supp}). The average nodal point energies $E_D = -5 \pm 1$ meV, lie within the KI gap, and their distinguishing velocities in momentum space, $v_{\bar{X}} \approx 1150 \pm 40 \ \mbox{m/s}$ and $v_{\bar{\Gamma}} \approx 14000 \pm 1000 \ \mbox{m/s}$, identify them as Dirac states positioned at the two symmetry-equivalent $\bar{X}$-points and the $\bar{\Gamma}$-point, respectively (Fig.\ \ref{F1}D).  The largest effective Dirac mass is at the $\bar{X}$ point with $m^* = \hbar k_F/v_{D_{\bar{X}}} = (410\pm20)m_e$. We note that $k_F$, the size of our Fermi pockets along $\mathbf{q}_y$, is in excellent agreement with ARPES measurements \cite{Jiang2013,Xu2014}. However, the apparent discrepancies between our measured velocities and Dirac points, and those of ARPES, can be explained by the fact that we access a single, non-polar surface, whereas ARPES experiments typically average over a mixture of terminations. The Dirac cones we image are in excellent agreement with expectations for a TKI phase with $f$-dominated heavy Dirac fermions \cite{Takimoto2011,Alexandrov2013,Lu2013}.


\mypara
Intra-cone backscattering is the simplest identifiable process responsible for the observed Dirac-state QPI.  We adduce that the scattering vector $\mathbf{q}(E_F) = 2\mathbf{k}(E_F)$ is in good agreement with the size of surface state Fermi pockets in electronic structure calculations \cite{Lu2013}. Although backscattering of topological surface states is generally suppressed in topological insulators, it has been observed to generate a QPI signal when the scattering involves a spin-flip in the plane, attributed to a magnetic scattering potential \cite{Okada2011}. To assess the magnetic character of defects in the Sm layer around which we detect QPI, we measured magnetic susceptibility and found that a small inclusion of nominally non-magnetic Sm-vacancies led to additional magnetic structure at low temperatures, akin to the addition of Kondo holes \cite{Figgins2011} (see section VIII of \cite{Supp}). 


\mypara
The formation of heavy Dirac fermion surface states is predicated on the coherence of the correlation-driven gap.  With increasing temperature, incoherent Kondo lattice states spread into the gap as the hybridization between $f$ and $d$ electrons unwinds. This process is expected to drive a topological phase transition at higher temperatures as band-parity-inversion is lost, eliminating the non-trivial surface states.  However, theory has not yet succeeded to describe coherent Kondo lattice evolution, much less the concomitant formation of heavy Dirac fermions. Here, we extract experimentally this complex connection by tracking the simultaneous temperature evolution of the KI gap and Dirac surface state contributions to $g(E)$. In general, $g(E)$ cannot be directly interpreted as density of states in multi-orbital Kondo lattice systems because of quantum interference between electrons co-tunneling into $f$ and $d$ states \cite{Maltseva2009,Figgins2010a}. We account for this effect by modeling the differential conductance spectrum as

\begin{equation}\label{eq:didv}
	g(\mathbf{r}_0,E) \propto \left[t^{T}\textrm{Im}G(\mathbf{r}-\mathbf{r}_0=0,E)t\right],
		\quad t^{T} = [t_d \; t_{f_1}\; t_{f_2}], 
\end{equation}
where $t_{\alpha}$ represent tunneling probabilities into the individual orbital states and $G(\mathbf{r},E)$ is the Fourier transform of $\widetilde{G}(\mathbf{k},E)$, the renormalized KI Green's function that implements a tight-binding Hamiltonian describing the known bulk bands (section III of \cite{Supp}).  Figure \ref{F4}A shows a fit to a single spectrum taken at 15K, demonstrating excellent parametrization of the $\smb$ KI electronic structure.  In the same panel, a fit at 2K reveals that the KI model, representing the bulk, does not capture the full set of states for energies within the gap, $\Delta$.  The deviations, plotted in Fig.\ \ref{F4}C for a sequence of temperatures, show a lineshape characteristic of linear bands with a crossing point near $-5$ meV, corroborating our independent QPI measurements of heavy Dirac surface states. Similar temperature-dependent features in $g(E)$ spectra have also been observed on the $(1\times1)$ boron termination layer \cite{Jiao2016}.
\mypara 

The emergence of surface states is connected to the development of the insulating gap, which is most clearly seen in the density of states for the $d$ and $f$ levels. The KI model provides a weighted sum of the density of states, $D(E)$, which can be used to track the bulk gap. Indeed, the temperature dependence of $D(E, T)$, presented in  Fig.\ \ref{F4}B, reveals a narrow energy window of diminished spectral weight that onsets below $T^*_{\Delta} \approx 35$ K and exhibits a gap of $\Delta \approx 10$ meV at low temperatures, in correspondence with bulk probes of $\smb$ \cite{Allen1979,Zhang2013,Kim2014,Syers2015}.  The surface state and in-gap spectral weights are presented in  Fig.\ \ref{F4}D and E, respectively. Foremost is their inverse relationship: the surface states diminish rapidly as the KI gap fills. We note that the Dirac weight diminishes faster than would be expected by thermal broadening alone (blue line in Fig.\ \ref{F4}D), and indeed faster even than the filling of the insulating gap around 35K where the topology of the bands is inverted.   One possible explanation for this fast decay is that the surface states are reliant on bulk coherence. Alternatively, Kondo lattice defects can generate magnetic correlations that suppress the onset of surface states to lower temperature \cite{Kim2014,Akintola2017}. 



\mypara
The visualization of correlation-driven topological surface states, a pressing challenge in recent years, has now been achieved by STM.  Simultaneous imaging of Kondo insulator formation and slow Dirac surface modes within the bulk gap provides direct evidence that $\smb$ is a topological Kondo insulator harboring the heaviest known Dirac fermions.  The optimal positioning of the $f$-character surface states at the chemical potential, enforced by Kondo lattice interactions, increases prospects for interface engineering to discover novel forms of topological superconductivity and construct transformative quantum devices.  $\smb$ and prospective TKIs may become leading testbeds for fractional and non-Abelian statistics, both of which are essential elements of prospective universal topological quantum computation. 

\vspace{1cm}

\textbf{Acknowledgements}
    The work at Harvard was supported by the U.S.\ National Science Foundation under grants DMR-1106023 and DMR-1410480.  The work at UC Irvine was supported by U.S.\ National Science Foundation under grant 1708199. D.K.M.\ acknowledges support by the U.S.\ Department of Energy, Office of Science, Basic Energy Sciences, under Award No.\ DE-FG02-05ER46225.
\vspace{0.11cm}

\textbf{Author Contributions} 
    H.P., Y.L., A.S., P.C., Y.H., and M.M.Y.\ performed STM experiments. X.W., J.P.P., P.F.S.R., D.J.K., and Z.F.\ synthesized and characterized the samples. P.F.S.R.\ performed X-ray measurement. J.D.T.\ performed magnetic susceptibility measurements. H.P., A.S., Y.H., M.M.Y., J.E.H., and M.H.H.\ developed and carried out analyses. D.K.M.\ provided theoretical guidance. J.E.H.\ and M.H.H.\ supervised the project.  M.H.H.\ wrote the paper with key contributions from H.P., D.K.M., and J.E.H. The manuscript reflects the contributions and ideas of all authors.

\textbf{Author Information}
    The authors declare that they have no competing financial interests.

Correspondence and requests for materials should be addressed to J.E.H. (jhoffman@physics.harvard.edu) and M.H.H. (m.hamidian@gmail.com)

\bibliography{ref}

\clearpage
\onecolumngrid
\begin{center}
\vspace{2cm}
\textbf{\large Supplementary Materials}
\vspace{0.5cm}
\end{center}
\twocolumngrid
\setcounter{equation}{0}
\setcounter{figure}{0}
\setcounter{table}{0}
\setcounter{page}{1}
\makeatletter
\renewcommand{\theequation}{S\arabic{equation}}
\renewcommand{\figurename}{{\bf Figure S}}
\renewcommand{\tablename}{{\bf Table S}}


\begin{figure}[h!] 
\includegraphics[width=0.5\textwidth]{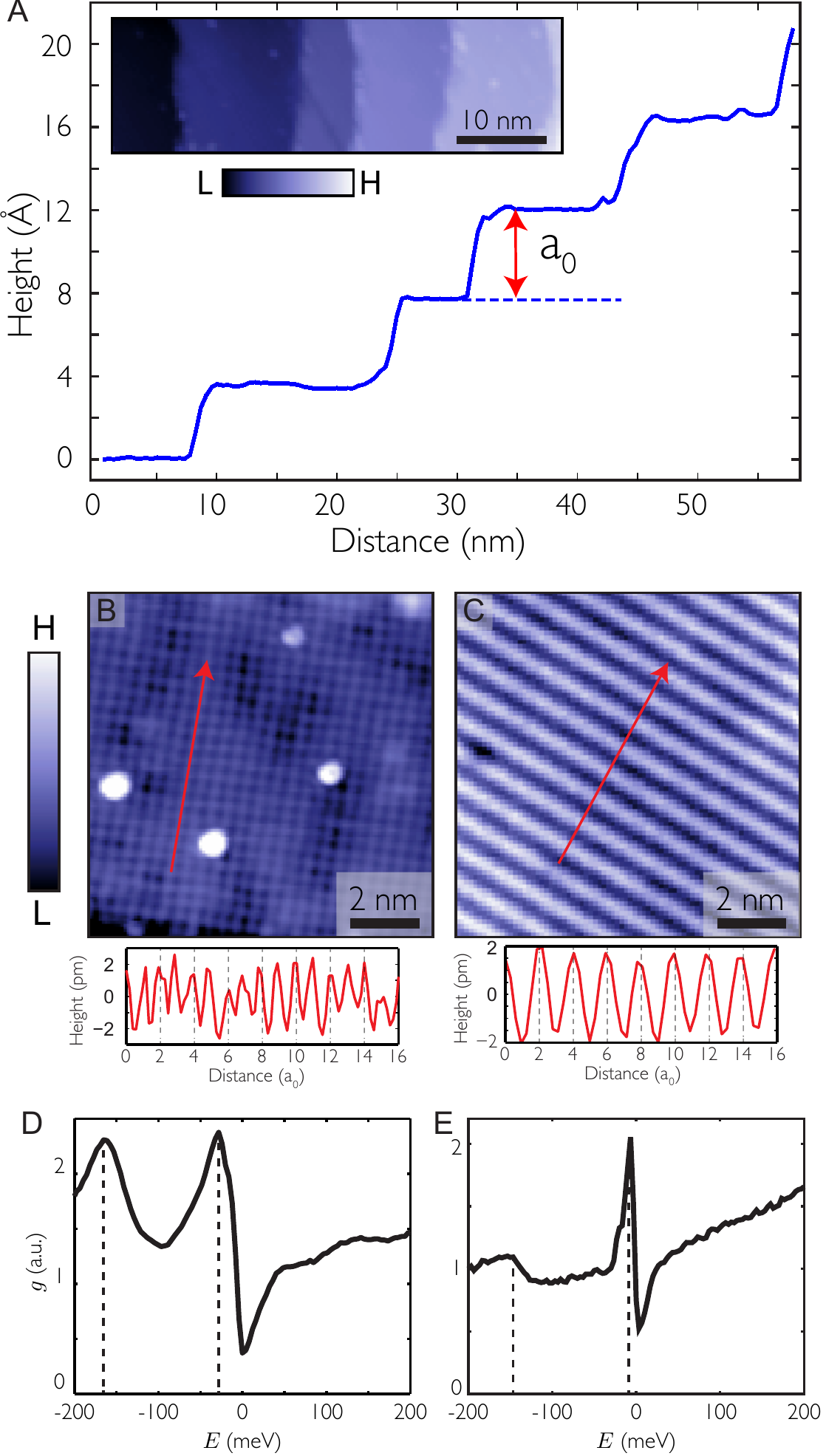}
\caption{ {\bf Typical surface morphology of $\smb$.}
    ({\bf A}) 
        Topographic line cut across five atomically flat terraces. The difference in the vertical height between adjacent terraces is $a_0$.  Inset shows a 50 nm $\times$ 15 nm topography of these terraces. ($V_s=-100$ mV, $R_J=10$ G$\Omega$.)
    ({\bf B})
        Representative images of $1\times1$ polar Sm termination ($V_s = -200$ mV, $R_J = 10$ G$\Omega$). Lower panel shows surface corrugation along the red arrow.
    ({\bf C})
        Representative image of  $2\times1$ non-polar, half-Sm termination ($V_s = -100$ mV, $R_J = 5$ G$\Omega$). Lower panel shows surface corrugation along the red arrow.
    ({\bf D-E}) 
        Spectral features in STM differential conductance $g(E)$ are shifted to lower energies on the $1\times1$ termination (D) than the $2\times1$ termination (E). 
    } \label{Fmorph}
\end{figure}

\begin{figure*} 
\includegraphics[width=0.85\textwidth]{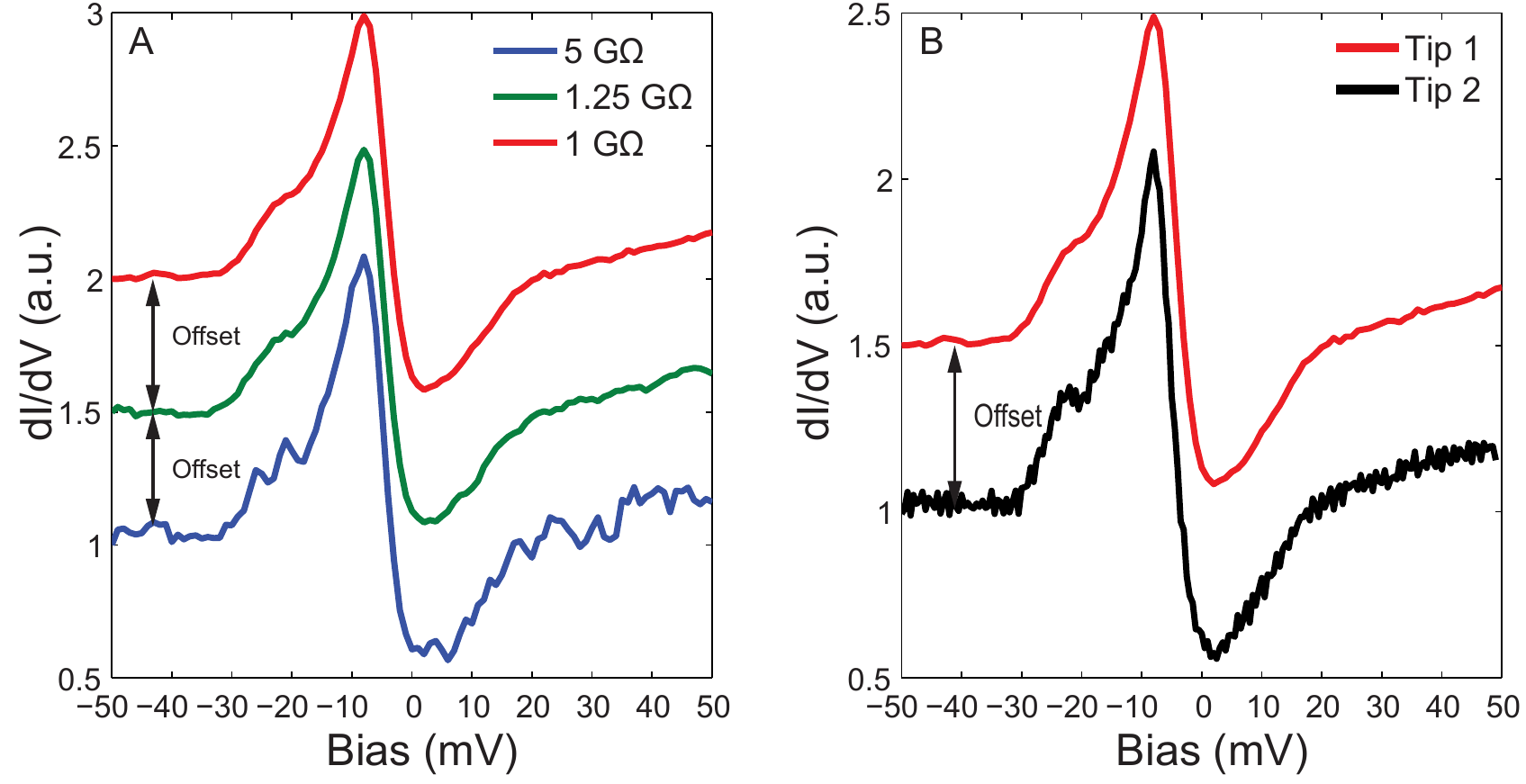}
\caption{ {\bf Insignificant tip-induced band bending on $\smb$.}
    ({\bf A}) 
        Variation of differential conduction spectra at different tip-sample separations. The junction resistance $R_\mathrm{J} = V_{\mathrm{s}}/I_{\mathrm{s}}$, where $V_{\mathrm{s}}$ and $I_{\mathrm{s}}$ are the spectroscopic bias and current feedback parameters, scales exponentially with the tip-sample separation for fixed bias. The spectra have been scaled by a multiplicative constant and offset from the $R_\mathrm{J}=$ 5 G$\Omega$ spectrum for clarity. All spectra were acquired at the same spatial location, equivalent acquisition parameters, and a common $V_{\mathrm{s}} = -100$ mV.  There is no apparent change in the shape of the spectra with tip-sample separation  indicating that the electric field of the tip in the experimental parameter regime does impact the position of the electronic bands.
    ({\bf B})
        Typical differential conductance spectra acquired with two different STM tips showing nearly identical features. The reproducibility of the spectra ensures consistent interpretation of data across many experiments even with tip variations.
    }\label{Fbandb}
\end{figure*}

\begin{figure*} 
\includegraphics[width=0.85\textwidth]{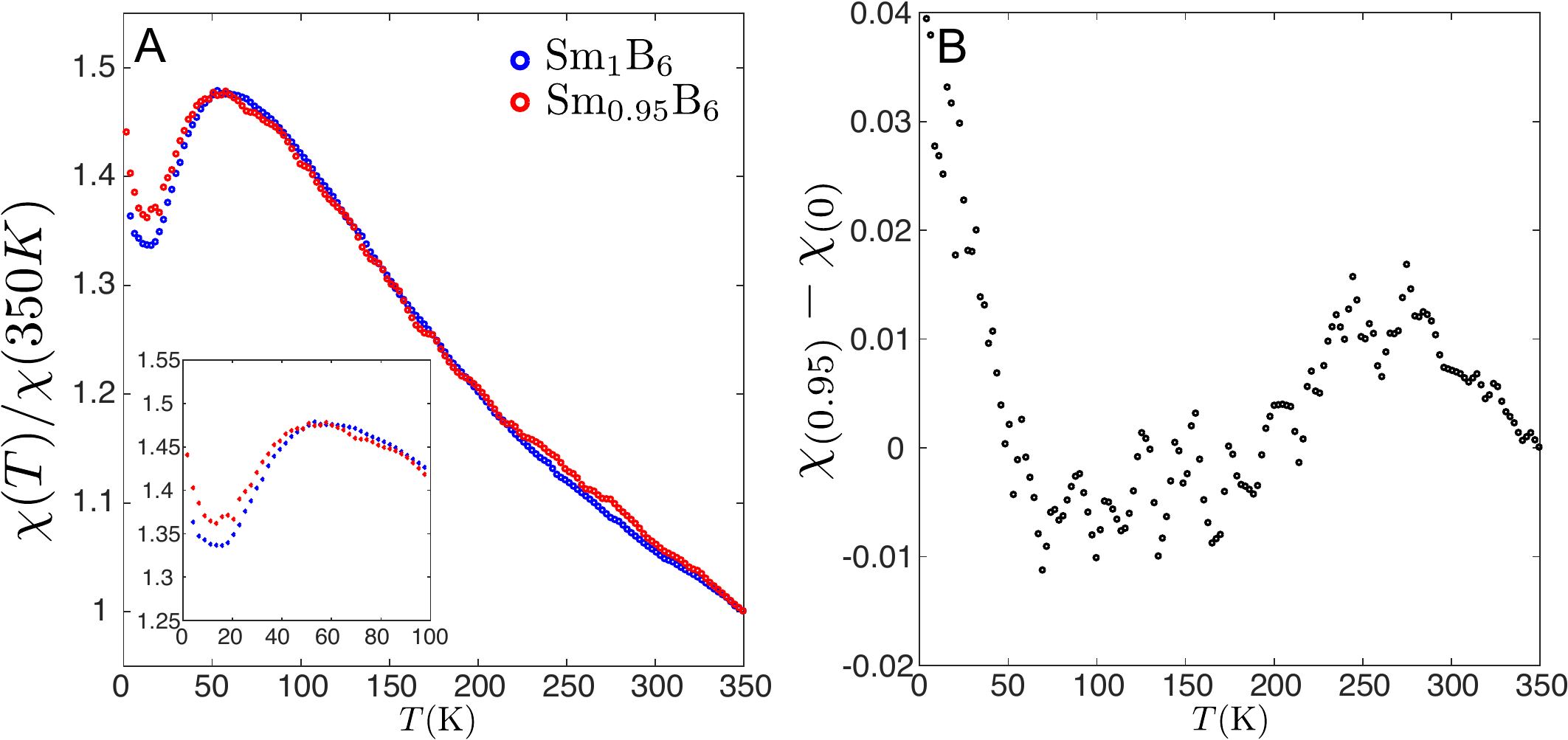}
\caption{{\bf Magnetic susceptibility increases with the addition of Kondo holes}
    ({\bf A}) 
        Temperature dependent magnetic susceptibility measurements comparing pure $\smb$ and Sm$_{0.95}$B$_6$.  The data has been normalized to their respective values at 350K. Below 60K, a relative increase in the susceptibility is observed in SmB$_{0.95}$.  
    ({\bf B}) 
        Temperature dependent difference in the normalized magnetic susceptibilities of $\smb$ and SmB$_{0.95}$.  Below 60K there is a divergence suggesting an increase in magnetic content in Sm$_{0.95}$B$_6$ with respect to $\smb$.  As the onset temperature is close to the Kondo lattice coherence temperature, the upturn in susceptibility is likely attributed to Kondo hole formation at the Sm vacancy sites. 
    }\label{Fsusc}
\end{figure*}

\begin{figure*} 
\includegraphics[width=0.8\textwidth]{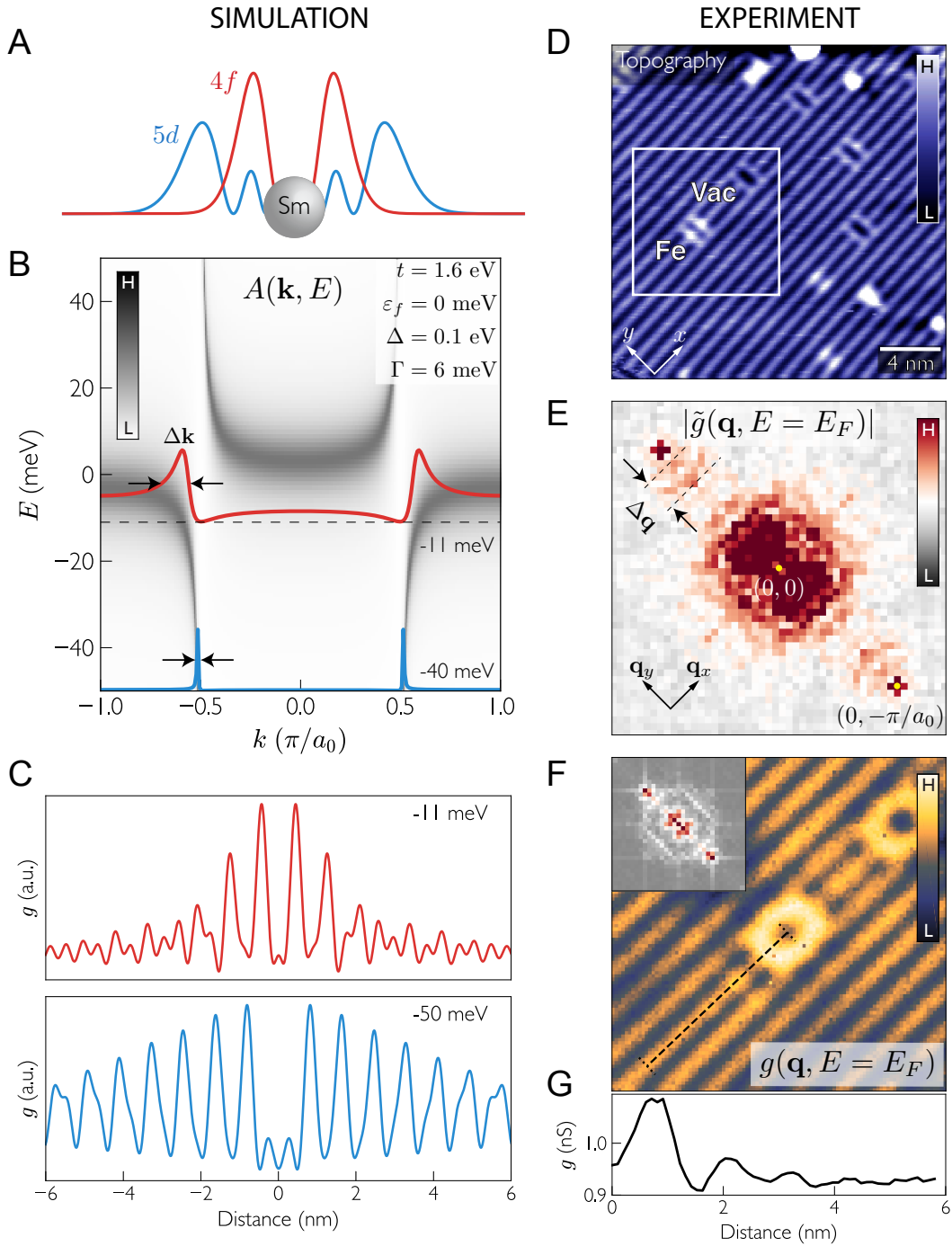}
\caption{ {\bf Heavy quasiparticles induce short-ranged interference patterns.}
    ({\bf A}) 
        Quasiparticles in $\smb$ contain contributions from both localized $f$ (red) and itinerant $d$ electrons (blue), depicted by the radial part of their wavefunctions. 
    ({\bf B})
        When the $d$ and $f$ bands hybridize, the effective mass of quasiparticles is renormalized, and they become non-dispersive close to the chemical potential (red trace), but dispersive away from it (blue trace)
    ({\bf C})
        The in-plane spatial decay length of the quasiparticles is determined by the $\mathbf{k}$-space full width at half maximum shown in (B); the real-space decay is shorter when the effective mass is large, as these quasiparticles contain a large contribution of localized $f$ states.
    ({\bf D})
        This 22 nm area on Fe-SmB$_6$ contains several quasiparticle scattering centers including an Fe substitution and a Sm vacancy.
    ({\bf E})
        The broad peaks along $q_y$ in the Fourier-transformed differential conductance reflect interference patterns caused by scattering heavy quasiparticles. This image has been two-fold symmetrized to increase signal to noise. 
    ({\bf F})
        In real space, the quasiparticle interference is short ranged around defects, reflecting the contribution from localized $f$ states. The oscillations create a ring in the corresponding Fourier transform at the Fermi wavevector (inset).  This data was recorded in the marked area in (D). 
    ({\bf G})
        A linecut away from the Fe impurity shows induced oscillations that decay after about three periods, similar to the simulation of heavy quasiparticles in (C) (red trace). 
    }\label{Fshortr}
\end{figure*}

\begin{figure*} 
\includegraphics[width=0.99\textwidth]{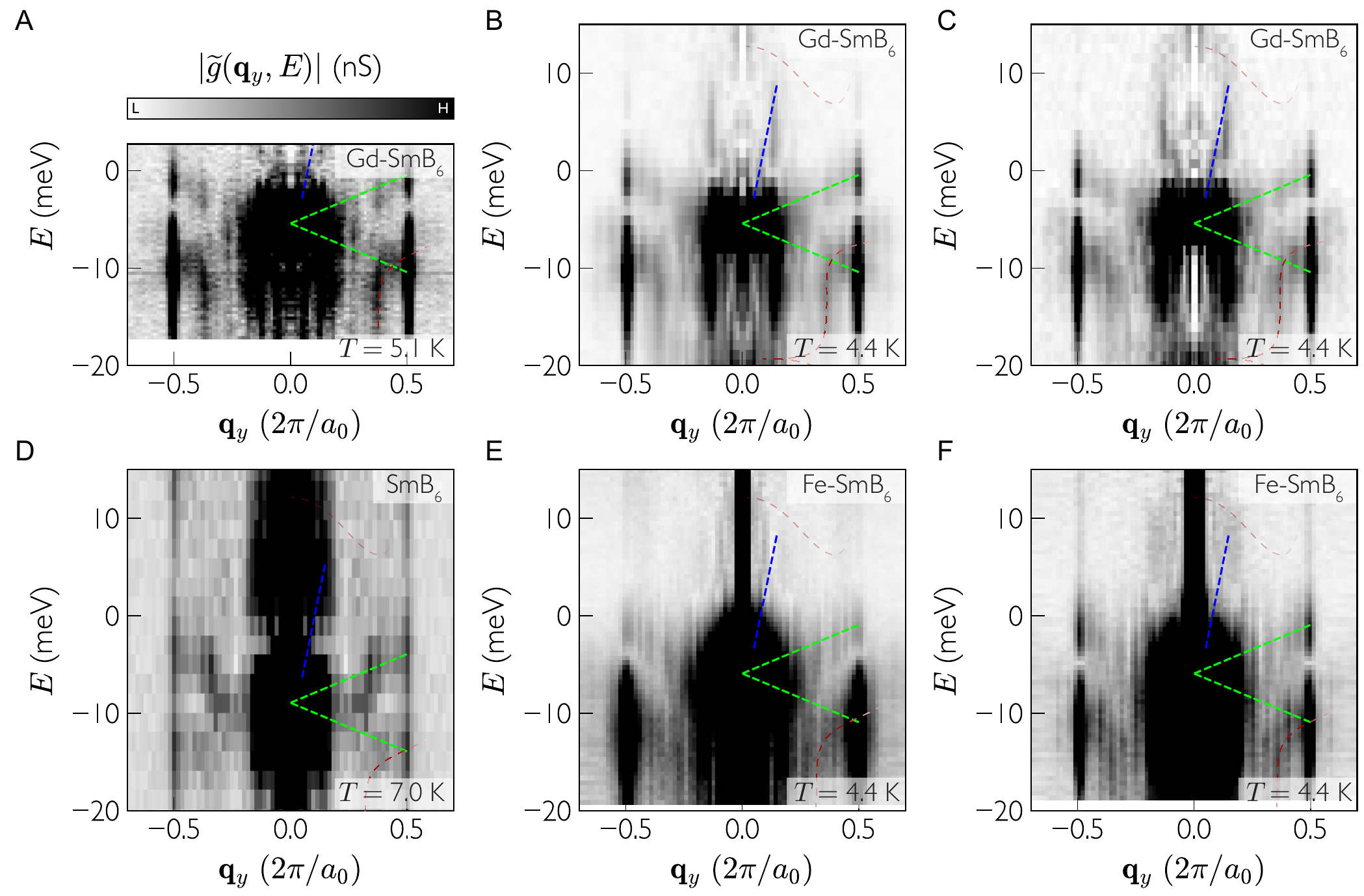}
\caption{{\bf Consistent QPI signatures in raw data from seven areas on three samples.}
    ({\bf A-C}) 
        Three different regions on the surface of Gd-doped $\smb$ give the same features in these arc-averaged linecuts of Hanning-windowed FT differential conductance, $|\tilde{g}(\mathbf{q},E)|$, along the $\mathbf{q}_y$-direction.  In each, quasiparticle interference contributes dispersing traces corresponding to scattering within three distinct bands:  bulk KI bands (red lines),  X-point Dirac states (green lines), and $\Gamma$-point Dirac states (blue line). 
    ({\bf D}) 
        The same qualitative QPI traces are measured on pure $\smb$ with lower resolution, and on ({\bf E-F}) Fe-doped $\smb$, where they  appear fainter, likely due to a higher doping level of 0.5\%.  Setup conditions for each data set are shown in table \ref{tbl:data}.
    }\label{FQPIcons}
\end{figure*}

\begin{figure*} 
\includegraphics[width=0.99\textwidth]{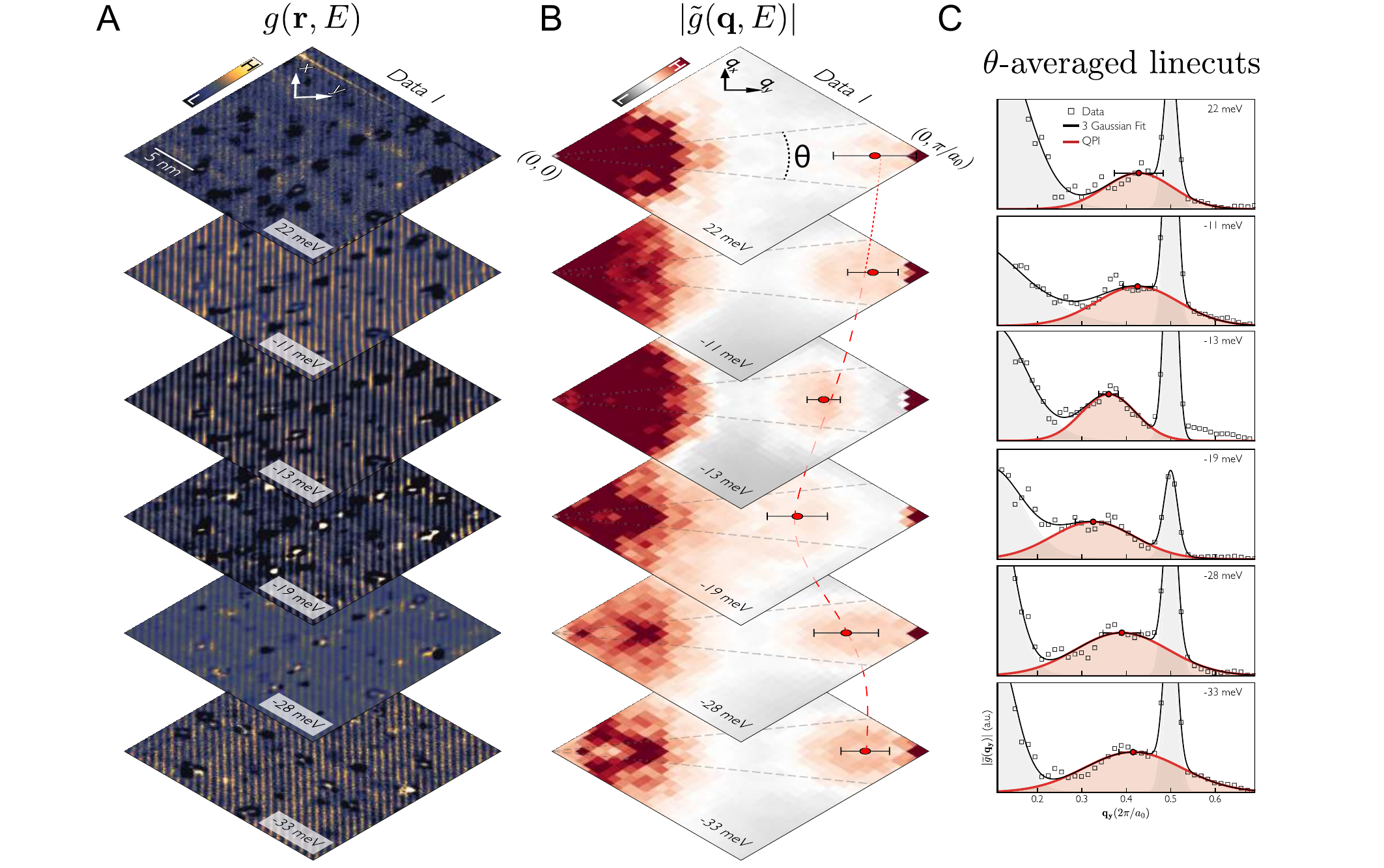}
\caption{{\bf  Quasiparticle interference of bulk KI states}
    ({\bf A}) 
        Spatially resolved maps of differential conductance, $g(\mathbf{r},E)$, shown at several energies on a $2\times 1$ reconstructed region (V$_\mathrm{setup} = 30$ mV, R$_J = 250$ M$\Omega$, lock-in amplifier modulation V$_\mathrm{rms}=1.5$ mV, $T=4.4$ K). 
    ({\bf B})
        Corresponding magnitude of their Fourier transforms $|\tilde{g}(\mathbf{q},E)|$, which were processed by two-fold symmetrizing along the $q_y$ diagonal, masking real-space defects to reduce low-$\mathbf{q}$ disorder, and with an edge-preserving Gaussian filter for clarity.
    ({\bf C})
        $\theta$-averaged linecuts of the unfiltered $|\tilde{g}(\mathbf{q},E)|$ along the $\mathbf{q}_y$ direction and their fits to a three-Gaussian model which captures dispersive peaks from quasiparticle interference (red Gaussian), as well low $\mathbf{q}$ disorder and the Bragg peak.
    }\label{FbulkQPI}
\end{figure*}

\begin{figure*} 
\includegraphics[width=0.99\textwidth]{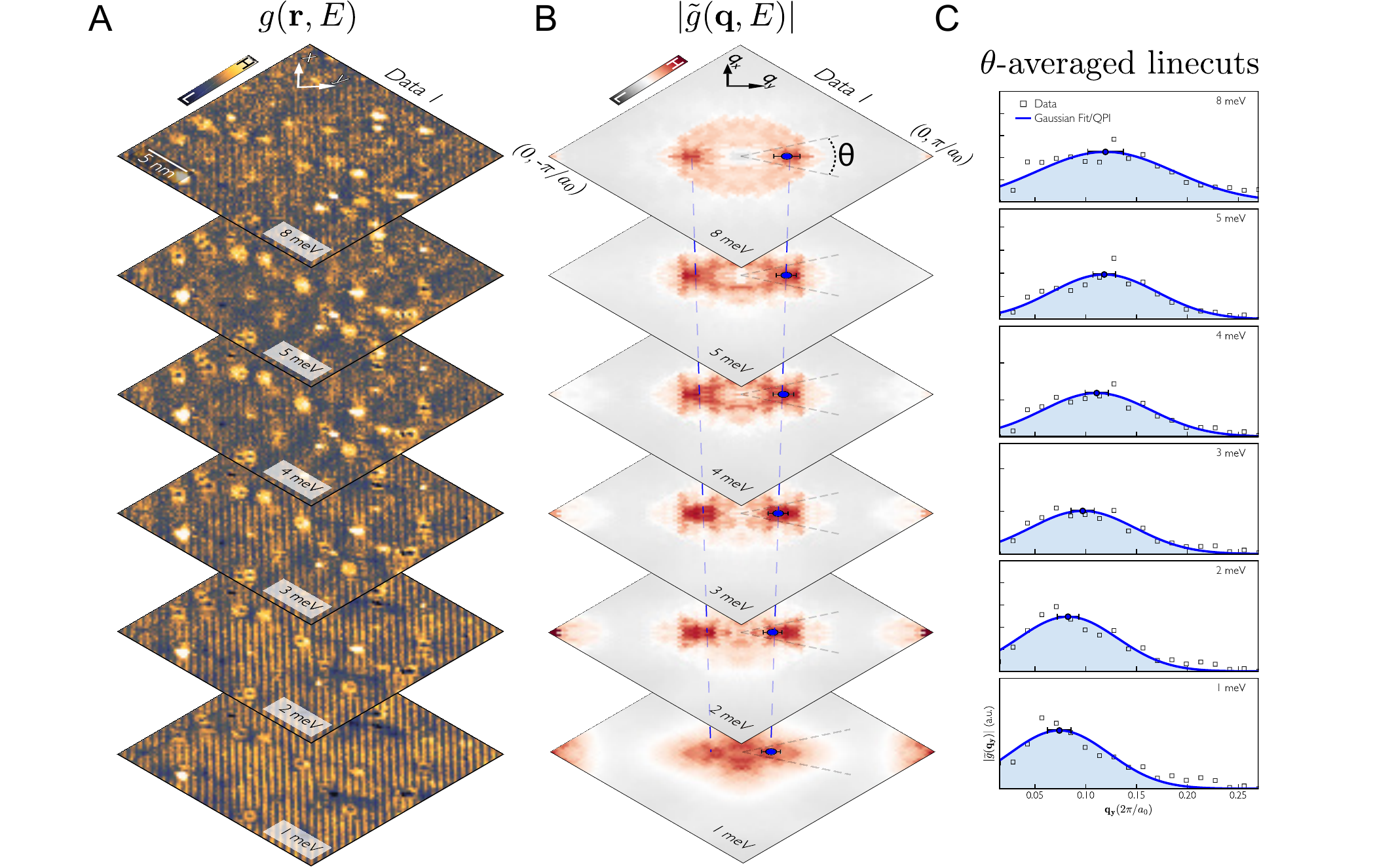}
\caption{{\bf  Quasiparticle interference of  $\Gamma$-point Dirac surface states}
    ({\bf A}) 
        Spatially resolved maps of differential conductance, $g(\mathbf{r},E)$, shown at several energies on a $2\times 1$ reconstructed region (V$_\mathrm{setup} = 30$ mV, R$_J = 250$ M$\Omega$, lock-in amplifier modulation V$_\mathrm{rms}=1.5$ mV, $T=4.4$ K). 
    ({\bf B})  
        Corresponding magnitude of their Fourier transforms $|\tilde{g}(\mathbf{q},E)|$, which were processed by two-fold symmetrizing along the $q_y$ diagonal, masking real-space defects to reduce low-$\mathbf{q}$ disorder, and with an edge-preserving Gaussian filter for clarity.
    ({\bf C})
        $\theta$-averaged linecuts of the unfiltered $|\tilde{g}(\mathbf{q},E)|$ along the $\mathbf{q}_y$ direction and their fits to a single Gaussian which captures the dispersion. 
    }\label{FGQPI}
\end{figure*}

\begin{figure*} 
\includegraphics[width=0.99\textwidth]{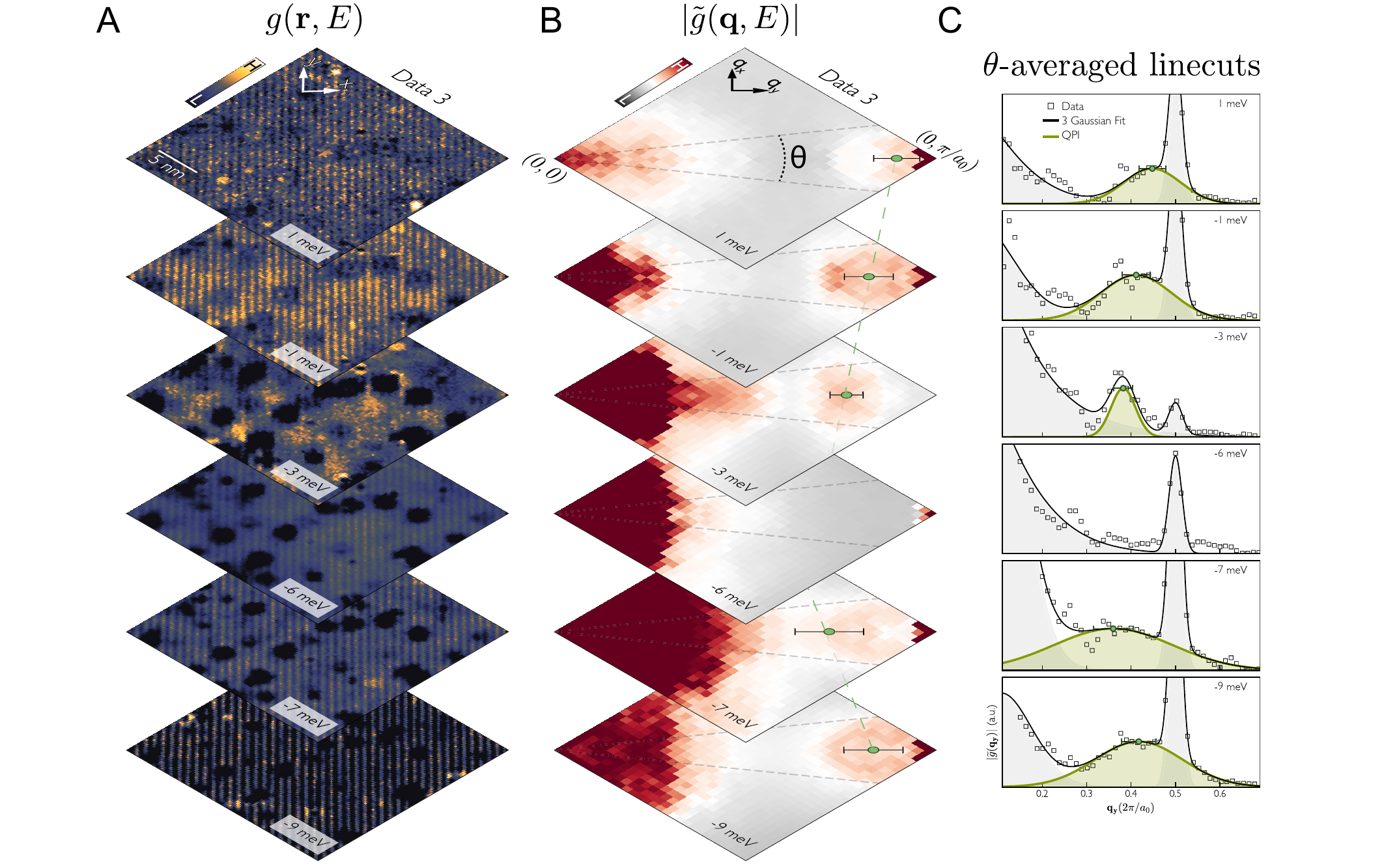}
\caption{ {\bf  Quasiparticle interference of $X$-point Dirac surface states}
    ({\bf A}) 
        Spatially resolved maps of differential conductance, $g(\mathbf{r},E)$, shown at several energies on a $2\times 1$ reconstructed region (V$_\mathrm{setup} = 30$ mV, R$_J = 170$ M$\Omega$, lock-in amplifier modulation V$_\mathrm{rms}=0.5$ mV, $T=5.1$ K). 
    ({\bf B})  
        Corresponding magnitude of their Fourier transforms $|\tilde{g}(\mathbf{q},E)|$, which were processed by two-fold symmetrizing along the $q_y$ diagonal, masking real-space defects to reduce low-$\mathbf{q}$ disorder, and with an edge-preserving Gaussian filter for clarity.
    ({\bf C})  
        $\theta$-averaged linecuts of the unfiltered $|\tilde{g}(\mathbf{q},E)|$ along the $\mathbf{q}_y$ direction and their fits to a three-Gaussian model which captures dispersive peaks from quasiparticle interference (green Gaussian), as well low $\mathbf{q}$ disorder and the Bragg peak.
    }\label{FXQPI}
\end{figure*}

\begin{figure*} 
\includegraphics[width=0.99\textwidth]{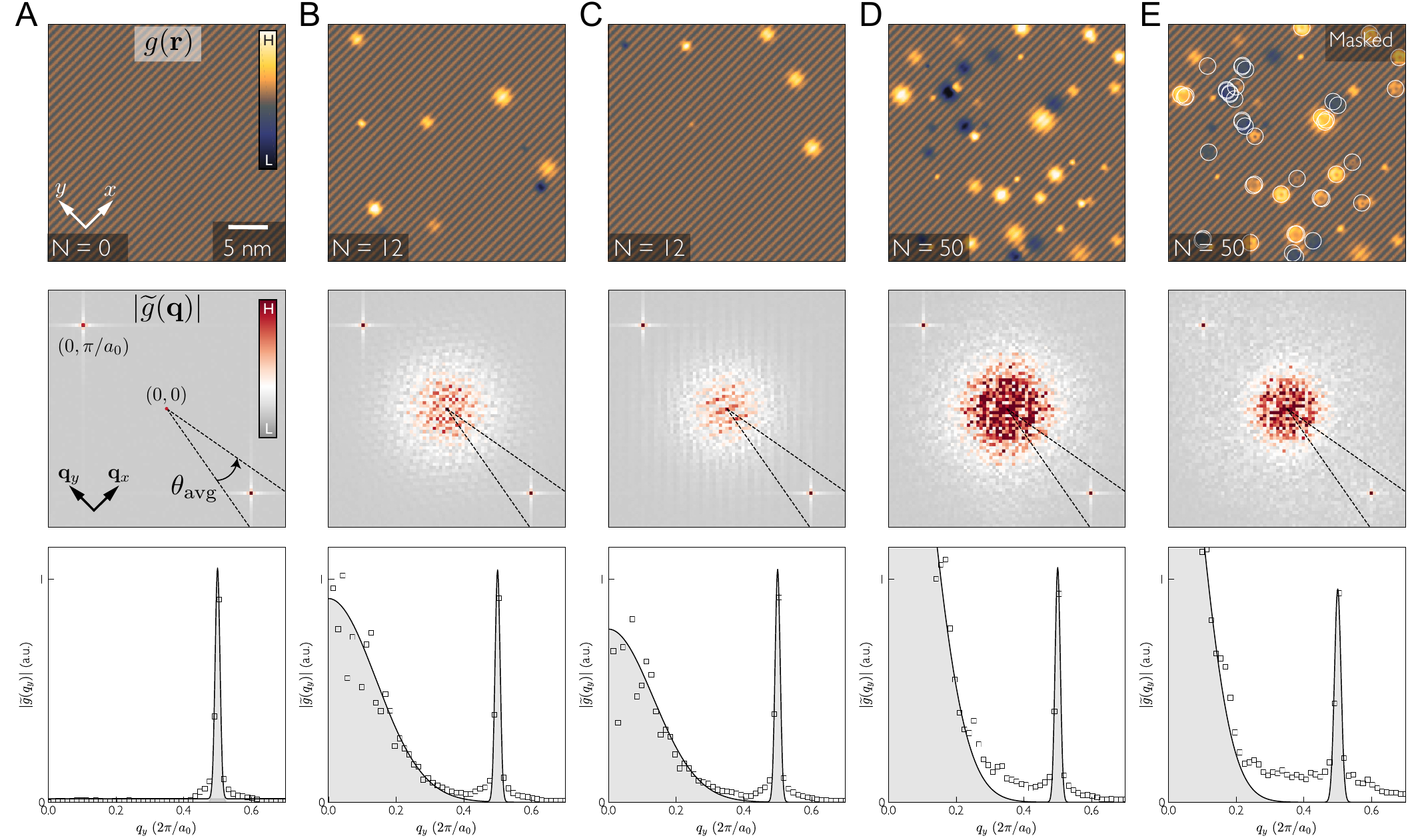}
\caption{{\bf Simulated random impurities create low-$\mathbf{q}$ disorder in FT.}
    ({\bf A}) 
        (top) Simulated contribution to differential conduction from the $(2\times 1)$ surface reconstruction in $\smb$.   (middle) The Fourier transform of panel (A) shows sharp peaks at the Bragg points, but no low-$\mathbf{q}$ disorder. (bottom) In a clean system all spectral weight in the arc-averaged linecut along $q_y$ appears around the Bragg peak with a slight leakage due to finite-size effects. 
    ({\bf B-C}) 
       Same as panel (A), but for a simulation that includes 12 randomly placed defects of random intensity and width.  Even a small number of defects induces low-$\mathbf{q}$ disorder in the FT (middle), which often produces ripples in the arc-averaged linecuts (bottom). 
    ({\bf D}) 
        Increasing the number of defects in the simulation to 50 greatly enhances low-$\mathbf{q}$ disorder. 
    ({\bf E}) 
        To reduce the effect of low-$\mathbf{q}$ disorder from random impurities, we apply a small Gaussian mask to the center of each defect before taking the Fourier transform.   
        }\label{Fdefect}
\end{figure*}

\begin{figure*} 
\includegraphics[width=0.99\textwidth]{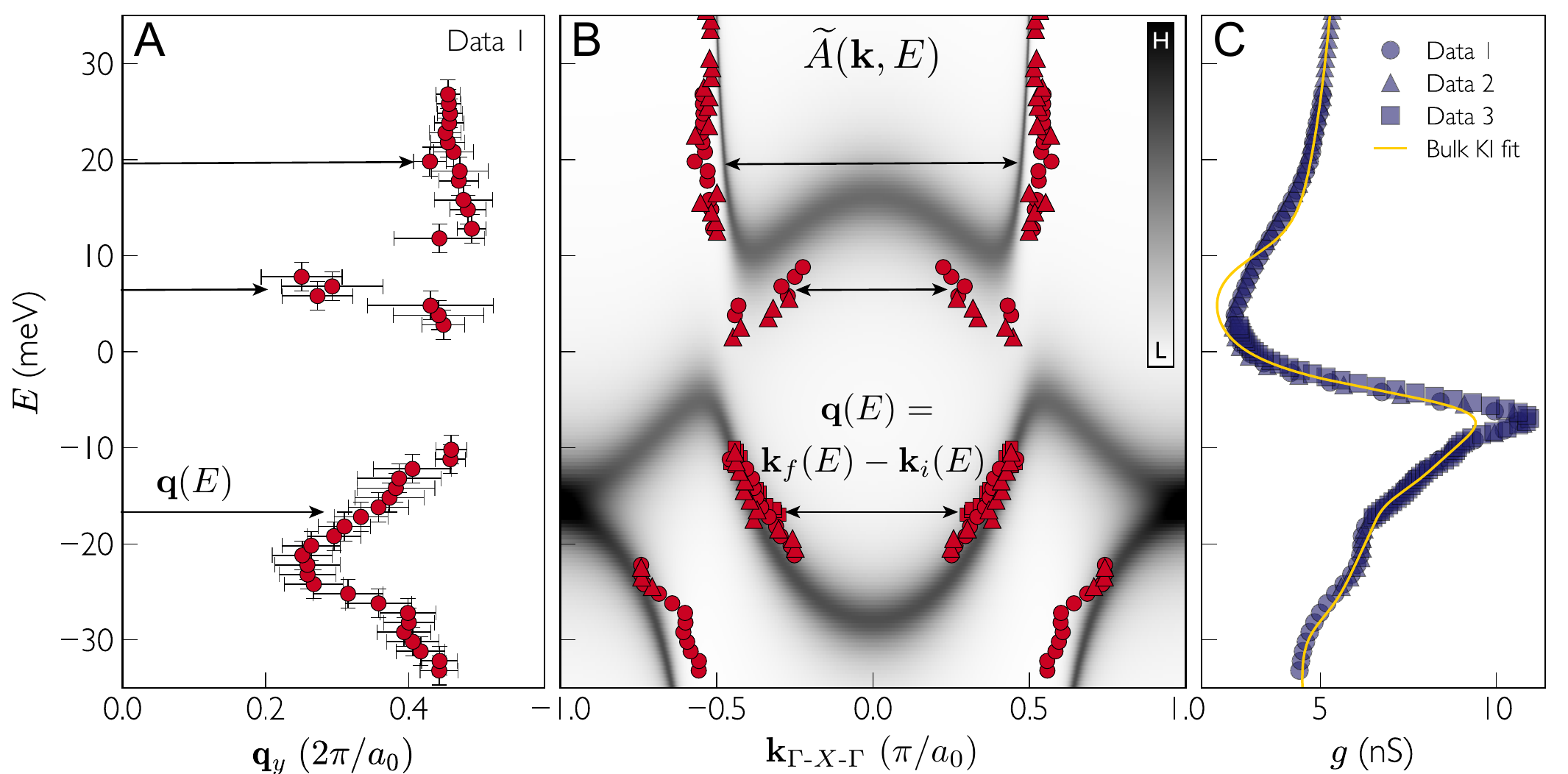}
\caption{{\bf Training a tight-binding model for the bulk KI states.}
    ({\bf A})
        Quasiparticle interference from bulk KI states creates intensity in Fourier transformed differential conductance measurements that follows this $\mathbf{q}(E)$ trajectory. The peak positions are determined from the fits shown in Fig.\ \ref{FbulkQPI}C, and are assigned to bulk KI states if the energy is outside the KI gap. 
    ({\bf B})
        At each such energy, $\mathbf{q}(E)$ measures the wavevector for elastic scattering of quasiparticle states, that is, it connects points on constant energy contours of the momentum-resolved spectral function, $\widetilde{A}(\mathbf{k}, E)$. 
    ({\bf C})
        Measured differential conductance, $g(E)$, fits well to the bulk KI model (equation (\ref{eq:KIdidv}), yellow line). Because the fit is calculated from a spectral function, as shown in (B), the KI model can be trained on both $g(E)$ and $\mathbf{q}(E)$ data: the KI model shown here is simultaneously fit to data sets from three distinct areas on the sample.  
    }\label{Ftrain}
\end{figure*}

\begin{figure*} 
\includegraphics[width=\textwidth]{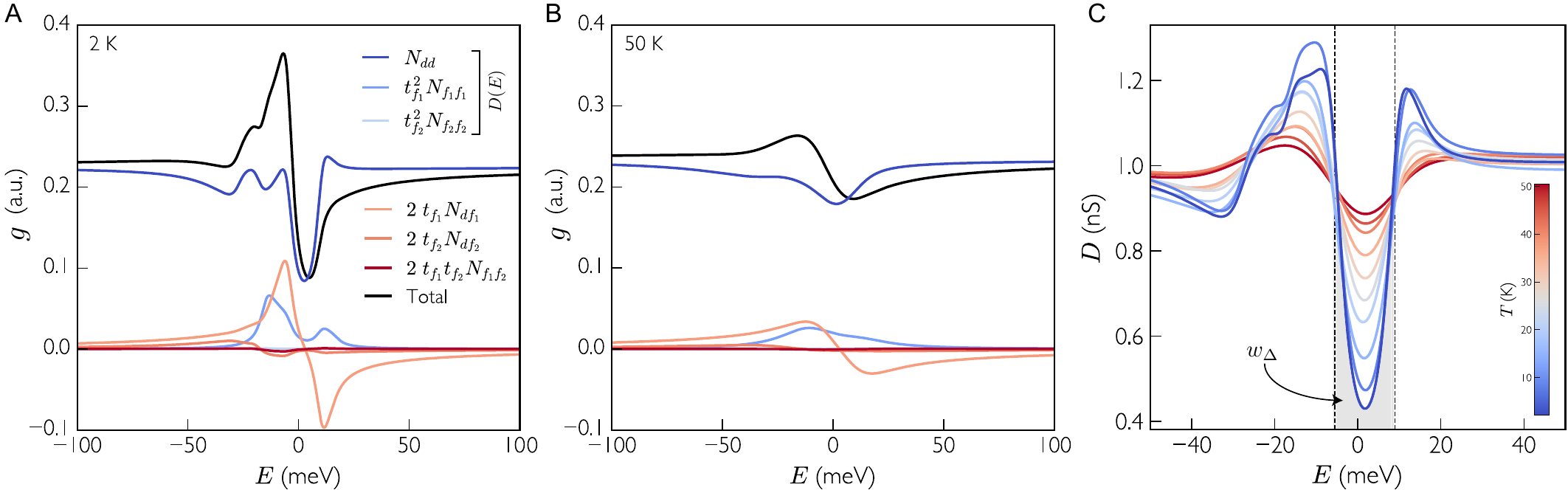}
\caption{{\bf Quantum interference between co-tunneling channels.}
    ({\bf A}) 
        Fitting $T$ = 2 K STM differential conductance spectrum to a Kondo insulator model gives contributions from three direct, $D$, channels (blue): $d$, $f_1$, $f_2$; and the three possible interference terms (red): $d$-$f_1$, $d$-$f_2$, $f_1$-$f_2$. The total spectrum (black) consists predominately of tunneling into the itinerant $d$ band, with a prominent peak-dip feature caused by $d$-$f_1$ interference.  The tunneling probability ratios have been normalized to $t_d = 1$.
    ({\bf B}) 
        Same decomposition as in (A), but for a fit to $T$ = 50 K data.
    ({\bf C})
        The temperature evolution of the direct bulk conductance $D (E)$ shows a hybridization gap gradually forming around the chemical potential as spectral weight $w_\Delta$ is pushed out. 
    } \label{Fqint}
\end{figure*}

\begin{figure} 
\includegraphics[width=0.5\textwidth]{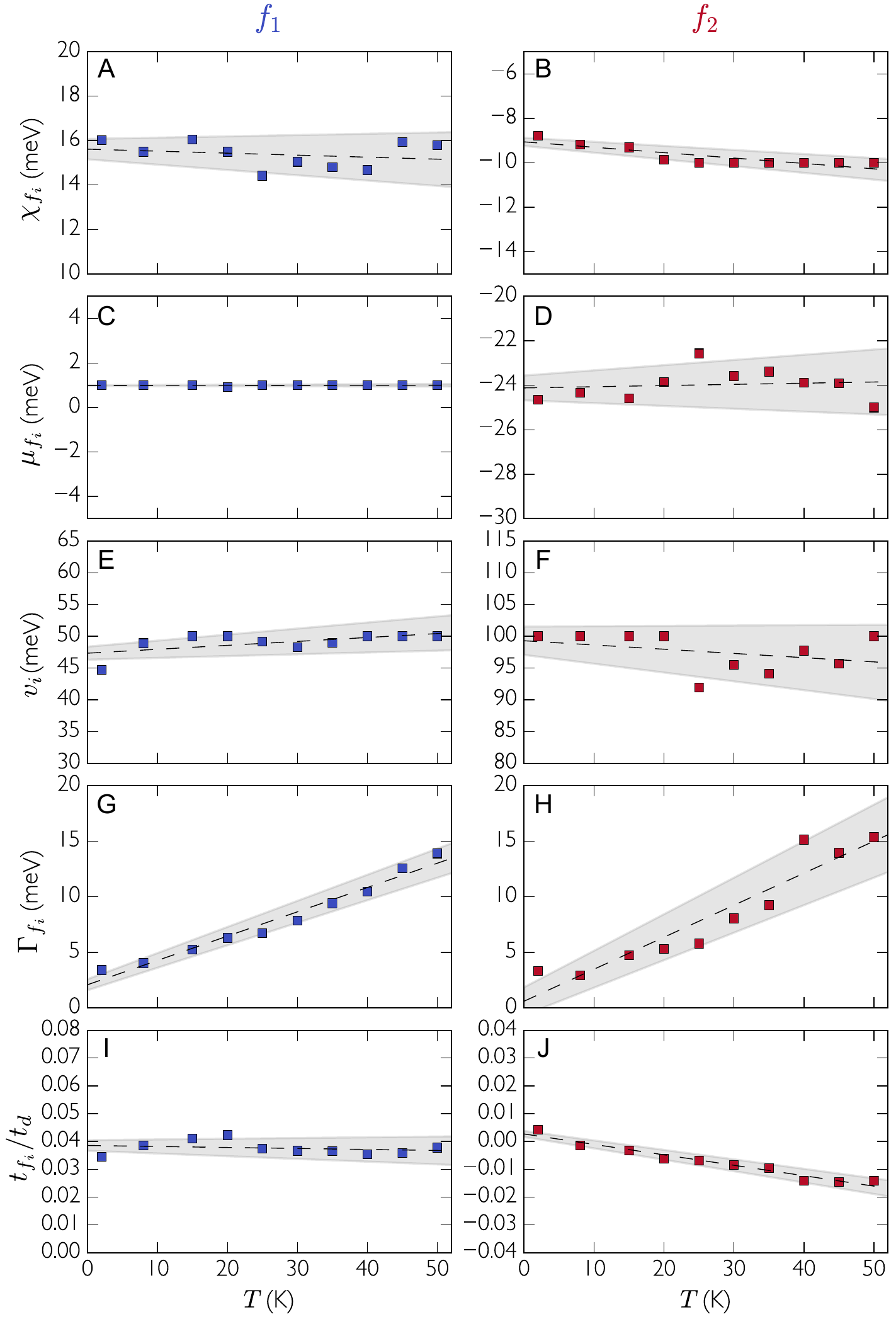}
\caption{ {\bf Temperature dependence of bulk KI electronic structure.}
    Temperature evolution of Kondo insulator electronic structure parameters involving the $f$ states. The parameters are evaluated by fitting STM differential conductance spectra to the Kondo insulator model in equation (\ref{eq:KIdidv}). The relevant parameters are shown for the upper $f_1$ level hybridization in blue and the lower $f_2$ level in red:  
    ({\bf A-B}) 
        $\chi_{f_i}$, half-bandwidth of $f$ states; 
    ({\bf C-D}) 
        $\mu_{f_i}$, middle of $f$ bands 
    ({\bf E-F}) 
        $v_i$, hybridization element between $f_i$ and $d$ states; 
    ({\bf G-H})
        $\Gamma_{f_i}$, inverse $f_i$ quasiparticle lifetime; 
    ({\bf I-J})
        $t_{f_i}/t_d$, relative tunneling probability between $f_i$ and $d$ channels.  The gray region is an estimation of the confidence interval of the linear trend-line.  The most significant temperature dependence is observed in the band broadening, $\Gamma_{f_i}$, indicating a loss of coherence with increasing temperature. 
    } \label{FKIfit}
\end{figure}

\begin{figure*} 
\includegraphics[width=0.9\textwidth]{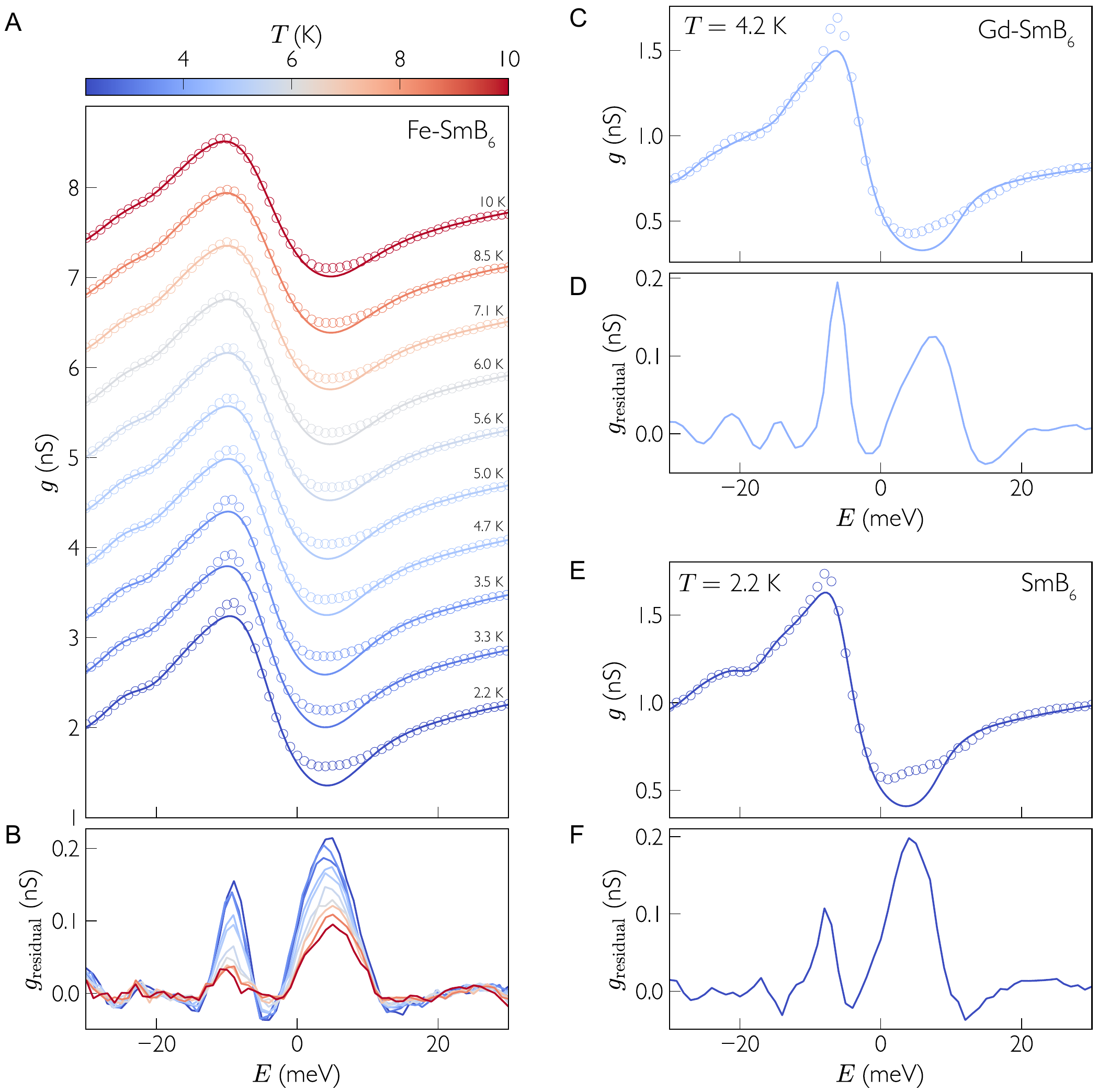}
\caption{{\bf Consistent temperature dependent Dirac surface states across multiple samples.}
        ({\bf A})  
            A fit to the Kondo insulator model in equation (\ref{eq:KIdidv}) (solid lines) captures the contribution from bulk-projected bands to differential conductance, $g (E)$ (open circles). However, as the temperature is lowered past 10 K, a significant deviation between the fit and the data develops for energies near the hybridization gap. 
        ({\bf B}) 
            Because the bulk contribution is removed by the fit, these deviations are attributed to residual surface conductance. Their V-shape is typical for a linear band dispersion, and contains a nodal energy of $-5$ meV that agrees with the Dirac point measured by QPI (Fig.\ 3C-D of the main text).  
        ({\bf C-F})
            Consistent low temperature deviations between $g(E)$ and the bulk KI model were measured on ({\bf C-D}) Gd-doped samples and  ({\bf E-F}) undoped samples.
        }\label{FSScons}
\end{figure*}

\begin{figure*} 
\includegraphics[width=0.81\textwidth]{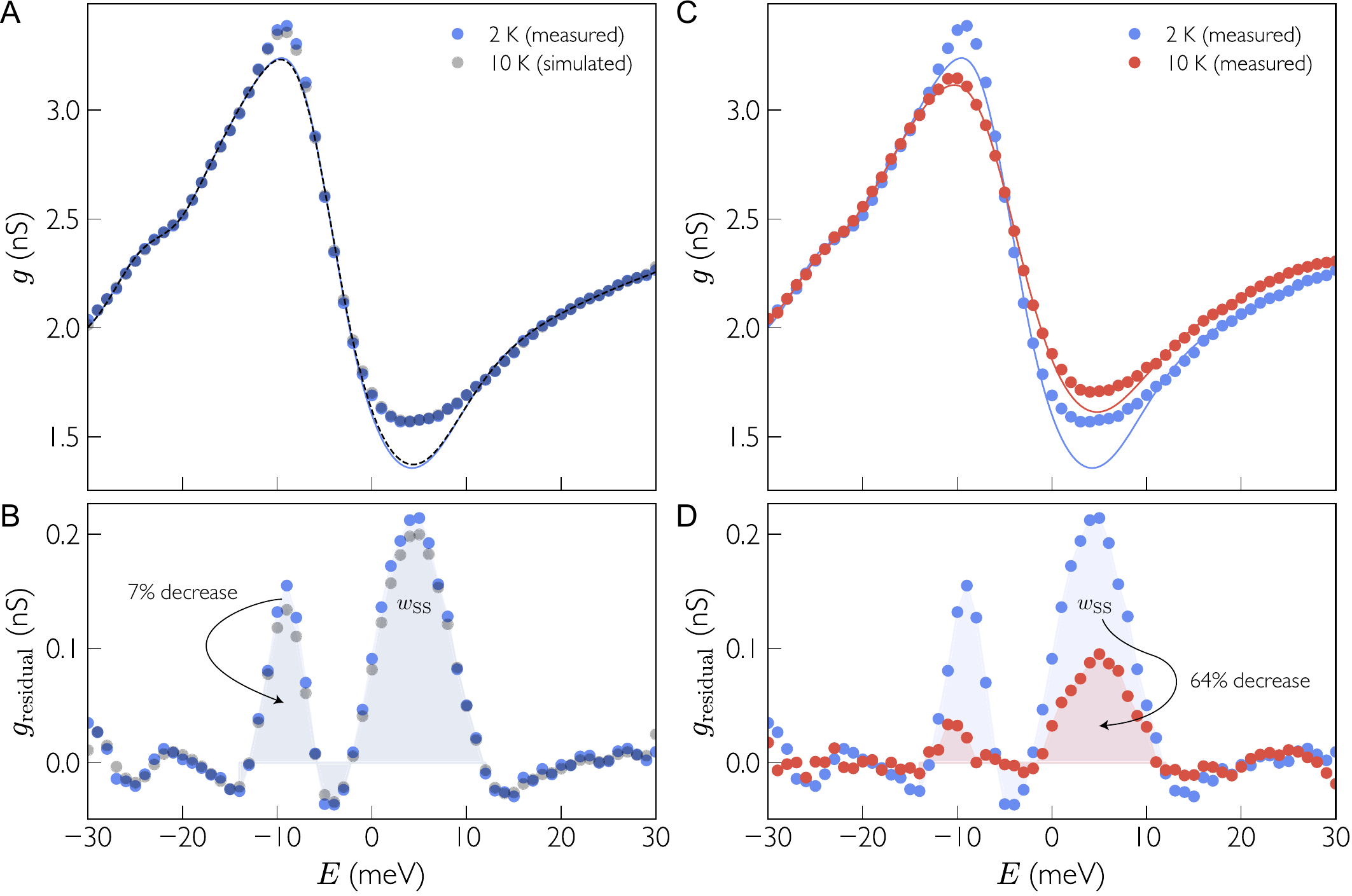}
\caption{{\bf Heavy Dirac states decay faster than predicted by thermal broadening.}
    ({\bf A}) 
        The effect of thermal broadening is simulated by convolving a measured $g(E)$ spectrum (blue circles) with the derivative of the Fermi-Dirac distribution evaluated at the target temperature (here 10K, black circles). Both spectra are then fit to the KI model in equation (\ref{eq:KIdidv}) (solid, dashed lines, respectively).
    ({\bf B})
        Both the 2 K measurement and the 10 K simulation show a pronounced deviation from their respective fits for energies around the KI gap, reflecting surface state contributions disregarded by the model.  In our simulation of thermal broadening, the integrated residual is predicted to decay by 7\% from 2 K to 10 K.  
    ({\bf C}) 
        Measured $g(E)$ spectra at 2 K (blue) and 10 K (red) and their respective fits to the KI model (solid lines).
    ({\bf D}) 
        The measured integrated residual decays much faster than expected: there is a 64\% change in the shaded areas, not the 7\% predicted in (B).  Here, the rapid decay is influenced by the entangled host KI insulator unwinding, which destroys the coherence of the topological ground state.
    }\label{FSSdecay}
\end{figure*}


\section{Materials and Methods}



  
Magnetic susceptibility measurements were performed on single crystalline samples of SmB$_{6}$ and Sm$_{0.95}$B$_{6}$ grown by the Al-flux technique with starting composition Sm:B:Al $=$ $1-x:6:700$ ($x = 0.0$ and $0.05$, respectively). The 
mixture of Samarium pieces, Boron powder ($99.99$\%) and Aluminum shots ($99.999$\%) was placed in an alumina crucible and loaded in a vertical tube furnace with ultra high-purity Ar 
flow. The furnace was heated to $1450^{\circ}$C for $12$~h followed by slow cooling to $1050^{\circ}$C at 2$^{\circ}$C/h. At  $1050^{\circ}$C the furnace was shut down, and the flux was removed at room temperature by etching with a NaOH solution. The atomic structure of the 
resulting single crystals was verified by X-ray diffraction at room temperature in a Bruker D8 Venture diffractometer using a Mo $K\alpha$ X-ray source. A Quantum Design superconducting quantum interference device was used to measure the magnetic response of the crystals to a magnetic field of $1$~kOe. The magnetic susceptibility curves were normalized to their values at $350$K because of the uncertainty in the determination of the actual number of Sm vacancies. 

STM experiments were carried out on single crystals of pure, 0.1$\%$ Gd-doped, and 0.5$\%$ Fe-doped $\smb$ grown using the Al-flux method \cite{Kim2013}.  Crystals were cleaved in cryogenic ultrahigh vacuum (UHV) at $\sim 30$K and immediately inserted into our home-built STM. STM tips were cut from PtIr wire and cleaned by \emph{in-situ} field emission on Au foil. The cryogenic UHV environment allowed the cleaved surface to stay clean for several months. Data were collected on three different samples and multiple fields of view.

\section{Supplementary Text}

\subsection{I. Sample characterization by STM}
The topographic image in Fig.\ S\ref{Fmorph} A shows the cleaved surface of SmB$_6$, with atomically flat terraces of typical 10-20 nm extent. These terraces are separated by steps of height equal to the cubic lattice constant $a_0 = 4.13$ \AA, which identifies the cleaved surface as the (001) plane. We observed several ordered and disordered surface morphologies, consistent with reports from other STM groups \cite{Ruan2014,Roßler2014}. Two of the ordered terminations are shown in Fig.\ S\ref{Fmorph} B and D. Interestingly, we measured spectroscopic signatures of two $f$-levels at lower energies on the $(1\times 1)$ surface than the $(2\times1)$ surface (compare Figs.\ S\ref{Fmorph} D and E), possibly due to band bending on a polar $(1\times 1)$ cleave. For this reason, the STM measurements in the text were carried out on a $(2\times1)$-reconstructed half-Sm termination, which avoids band bending and trivial surface states arising from a polar termination \cite{Hlawenka2018,Zhu2013}. 

\subsection{II. Insignificant tip-induced band bending in $\smb$}
Tip-induced band bending can introduce systematic errors in STM electronic structure studies of semiconductors \cite{Feenstra2007}.  Local modifications to the Fermi level in the material occur when some fraction of the applied potential between the sample and the probe tip is dropped in the sample itself. The energy scale of the change is dependent on the tip-sample distance, shape of the tip, applied voltage, and size of the semiconducting gap. Even at lowest temperatures the $\smb$ excitation gap is small, and the typical STM bias voltages used to probe spectral features are also small suggesting that tip-induced band bending is not a serious consideration. We verified this empirically by examining the STM differential conductance spectrum with tip-sample separations and different tips, as detailed in Fig.\ S\ref{Fbandb}.  The energy scales of all spectra are unaltered by the various experimental parameters, demonstrating that tip-induced band bending does not play a significant role in the analysis in this paper.

\subsection{III. Backscattering in strongly correlated topological materials}
Our measured QPI patterns of the topologically emergent surface states in $\smb$ are a consequence of backscattered or nearly backscattered electrons across the Dirac cones.  In theoretical models of topological insulators where the band and spin structure of the Dirac cones is described solely by a Rashba spin-orbit interaction, QPI arising from intra-cone backscattering (both for magnetic and non-magnetic defects) is suppressed by the helical spin structure of the cone \cite{Guo2010}.  However, any canting of the in-plane spin structure, for example, by a Dzyaloshinski-Moriya interaction, will immediately enhance the backscattering signature in the QPI spectrum.  Such canting can also occur due to local spin-polarizations induced by magnetic defects \cite{Nyberg2008}, or in topological insulators with strong interactions, such as TKI, where it was shown theoretically that surface states possess an out-of-plane spin polarization \cite{Efimkin2014, Baum2012}.  As it turns out, QPI patterns induced by backscattering of topological surface states have previously been detected in the magnetically doped topological insulator Bi$_{2-x}$Fe$_x$Te$_3$ \cite{Okada2011}.


In strongly correlated systems, defects that arise from the presence of nominally non-magnetic atoms, dislocations, or simply missing atoms, can still induce magnetic moments in their vicinity, such that the effective defect (bare defect plus induced magnetic moment) is magnetic in nature. This scenario has been observed in great detail in cuprate superconductors and theoretically resolved by accounting for the strong antiferromagnetic fluctuations present even in the paramagnetic phase \cite{Hoglund2007}. A similar effect was observed in the spin-gap compound CuGeO$_3$ \cite{Manabe1998}. In the heavy fermion compound CeCoIn$_5$, NMR measurements revealed that Cd dopant atoms nucleate antiferromagnetic islands \cite{Urbano2007}.  Indeed, scenarios in which nominally non-magnetic become effectively magnetic require only the presence of strong antiferromagnetic correlations, which are also present in the Kondo lattice of $\smb$.  

To resolve the magnetic nature of defects in our samples that induce responsible for QPI, we measured the magnetic susceptibility of nominally pure, stoichiometric $\smb$ and crystals grown with a controlled increase in the concentration of vacancies at the Sm sites (see Materials and Methods).  The results, presented in Fig.~S\ref{Fsusc}, show that defects at the Sm sites produce an enhancement of the susceptibility below the Kondo lattice coherence temperature of $\approx$60K.  Recently, measurements on 0.5$\%$ Fe-doped $\smb$ also showed a similar enhancement of the low temperature susceptibility \cite{Akintola2017}, indicating that defects in the Sm matrix manifest as magnetic objects at low temperatures.  Both observations are in reasonable agreement with earlier  studies of Kondo lattice compounds with substitutions at the rare-earth sites.  For example, in CePd$_3$ \cite{Lawrence1996}, small concentrations of La were substituted at the Ce sites and the enhanced low temperature magnetic susceptibility was shown to be consistent with Kondo hole formation at the La sites, yielding emergent magnetic defects. Thus, independent of whether the scatterers that give rise to the QPI signal are nominally magnetic or non-magnetic at high temperatures, the intrinsic defects in $\smb$ likely become magnetic at low temperatures allowing backscattering within the helical surface states that we observe in our QPI patterns.  As a consistency check, we also substituted 0.1$\%$ Gd at the Sm sites, which act as magnetic defects at all temperatures \cite{Kim2014}.  We found that the QPI response in the Gd-doped sample was similar to the undoped sample, where scattering occurs mainly around Sm vacancy sites.

 \subsection{IV. Consistent quasiparticle interference signal in multiple samples}
 \begin{table*}
 \caption{\label{tbl:data} Experimental conditions for differential conductance maps}
  \begin{ruledtabular}
\begin{tabular}[c]{@{}cccccccc@{}}
\toprule 
Data (Fig.~S\ref{FQPIcons}) & Sample & Area (nm) & $V_\mathrm{setup}$ (mV) & $R_J$ (G$\Omega$)  & $V_\mathrm{rms}$ & $T$ (K) & $\Delta E$ (meV)
\\ \hline
    A   & Gd-SmB$_6$ & $30\times30$  & 30   & 0.17   & 0.5   & 5.1     &  1.7
\\
    B   & Gd-SmB$_6$ & $30\times25$  & 30   & 0.25   & 1.5   & 4.4     &  2.6
\\
    C   & Gd-SmB$_6$ & $30\times30$  & 30   & 0.25   & 1.5   & 4.4     &  2.6
\\
    D   & SmB$_6$    & $20\times20$  & -100 & 1.0    & 3.0   & 7       &  5.1
\\
    E   & Fe-SmB$_6$ & $30\times30$  & 30  & 0.15    & 0.5   & 4.4     &  1.5
\\
    F   & Fe-SmB$_6$ & $30\times30$  & 30  & 0.15    & 0.5   & 4.4     &  1.5
\\
\bottomrule
\end{tabular}
 \end{ruledtabular}
\end{table*}

 We conducted spectroscopic-imaging STM experiments on non-polar, $(2\times1)$ Sm-surface terminations to measure the Kondo insulator and emergent Dirac surface state electronic structure. On these surfaces, quasiparticle interference (QPI) around defects generates short-ranged spatial modulations in STM differential conductance $g(\mathbf{r}, E)$, which correspond to broad peaks in the Fourier transform (FT) of $g(\mathbf{r}, E)$.  Our QPI signal is attributed to three distinct states: $\bar{X}$-point heavy Dirac fermions, $\bar{\Gamma}$-point Dirac fermions and heavy fermions from the bulk Kondo insulator.  Each of these states has a high renormalized mass, due to a contribution from localized $f$ electrons.  Consequently, the interference pattern induced by heavy quasiparticles is expected to be short range, inheriting the localized nature of the $f$ electrons.  Mathematically, a short-range QPI pattern is characterized by broad features in $\mathbf{q}$ space; the full width at half maximum (FWHM) reflects the in-plane decay length of the QPI envelope.  In other words, as quasiparticle bands become heavier, constant-energy cuts through the spectral function yield peaks with a larger FWHM, corresponding to more localized states inducing shorter-range QPI patterns. We illustrate this effect for a simple Kondo insulator model that includes a dispersive $d$ band and a non-dispersive $f$ level, described by the Hamiltonian 
 \begin{align}
     \mathcal{H} =  \sum_k 2t\cos(ka) d^\dagger_k d_k 
        + \varepsilon_f f^\dagger_k f_k 
        - \Delta( d^\dagger_k f_k + f^\dagger_k d_k).
 \end{align}
  The spectral function is calculated from the standard definition
 \begin{align}
    \label{eq:green1}
    \tilde{A}(\mathbf{k},E) &= -\mbox{Im}\left[(E+i\Gamma) \mathbbm{1}-\mathcal{H}(\mathbf{k})\right]^{-1}/\pi,
\end{align}
where $\Gamma$ is the inverse quasiparticle lifetime and $\mathbbm{1}$ is the $2\times2$ identity matrix. The bands in this model flatten near the energy of the localized state, causing peaks in the spectral function to broaden, as shown in Fig.~S\ref{Fshortr}B. The possible scattering vectors are described by the joint density of states (JDOS), calculated from the autocorrelation of the spectral function. As expected, the real-space QPI signal, calculated from the inverse FT of the JDOS, decays rapidly for heavy quasiparticle states (red trace in Fig.~S\ref{Fshortr}C), but remains long lived for itinerant quasiparticles (blue trace in Fig.~S\ref{Fshortr}C).  Experimentally, QPI generates broad features in $\mathbf{q}$ space, with a width at the Fermi level of $\Delta \mathbf{q} \approx 0.2\ \pi/a_0$, as shown in Fig.~S\ref{Fshortr}E.  Correspondingly, the real-space QPI pattern decays after approximately three periods (Fig.~S\ref{Fshortr}F-G). 

The short-ranged nature of our QPI signal yields broad features in $\mathbf{q}$ space that are most readily visible in the raw linecuts along $q_y$, as in Fig.~\ref{F2} of the main text. We measured consistent QPI patterns in six raw datasets from three samples (see Fig.~S\ref{FQPIcons}), over a range of experimental conditions listed in table \ref{tbl:data}. In most cases, the momentum resolution is inherently limited by the width of the QPI, rather than the experimental setup. Our maximum energy resolution is 1.5 meV, calculated as the convolution of the derivative of the Fermi function with the semi-circular kernel imposed by the bias modulation. We found that the signal is clearest in the Gd-doped sample, and noticeably diminished in the Fe-doped sample, likely because of the increased doping concentration in the latter case.  

For each of the main signals we detect, we show an extended subset of real-space differential conductance maps, their corresponding FT, and linecuts along the highest signal-to-noise direction $\mathbf{q}_y$, in Figs.\ S\ref{FbulkQPI}-\ref{FXQPI}.  To visualize the broad QPI signal in the images of FT differential conductance (panel B in Figs.\ S\ref{FbulkQPI}-\ref{FXQPI}) we took two further processing steps: we reduce the low-$\mathbf{q}$ noise by applying a small Gaussian mask around each defect before Fourier transforming, see Fig.~S\ref{Fdefect} and section V; and we apply an edge-preserving Gaussian filter that averages nearby pixels only if they have similar intensity \cite{Tomasi1998}. The latter step is unnecessary for the quantitative extraction of QPI peaks from the linecuts shown in panel C of Figs.\ S\ref{FbulkQPI}-\ref{FXQPI}, which were taken from the unfiltered data.  To increase the signal-to-noise ratio in these linecuts, we azimuthally averaged over a small fixed angle of $\sim \ang{20}$ about the predominant scattering direction, $\mathbf{q}_y$, for each energy.  

Our FT differential conductance measurements contain three main contributions of varying magnitude: low-$\mathbf{q}$ disorder, a $|\mathbf{Q}_{\mathrm{Bragg}}| =\pi/a_0$ Bragg peak for the 2$\times$1 surface reconstruction, and broad QPI peaks generated by elastic scattering of heavy quasiparticle states \cite{Schmidt2010,Aynajian2012,Allan2013}. At each energy, we disentangle these contributions by fitting the azimuthally averaged linecuts with a combination of three Gaussians, for the bulk KI states (Fig.\ S\ref{FbulkQPI}C) and the $\bar{X}$-centered Dirac states (Fig.\ S\ref{FXQPI}C); or one Gaussian for the $\bar{\Gamma}$-centered Dirac states  (Fig.\ S\ref{FGQPI}C), which are far away from the Bragg peak and in an energy range where the effect of disorder is small.  By isolating the $\mathbf{q}(E)$ trajectory of the QPI peaks, we reconstruct the $\mathbf{k}$-space dispersion of each band from $\mathbf{q}(E) = 2\mathbf{k} (E)$, the process valid for intra-cone back scattering (Fig. S\ref{Ftrain} and Fig.\ \ref{F2} in the main text).


\subsection{V. Impact of defects on quasiparticle interference}

QPI patterns, observed in the Fourier transform of spatially resolved $g(\mathbf{r},E)$, are generated by defects that scatter quasiparticle states. In Fig.~S\ref{Fdefect}, we simulate how a random distribution of featureless impurity states on an otherwise clean 2$\times$1 reconstructed $\smb$ surface can produce a modulated spectral weight in the Fourier transform at small $\mathbf{q}$. We find that even a small number of defects induces low-$\mathbf{q}$ disorder in the FT (Fig.~S\ref{Fdefect}B-C). The amount of low-$\mathbf{q}$ disorder can be reduced by applying a small Gaussian mask to the center of each defect before Fourier transforming (Fig.~S\ref{Fdefect}D-E).  This masking technique helps to quantitatively determine the quasiparticle dispersion, but the dispersion itself is qualitatively visible in the raw linecuts along $q_y$ (Fig.~S\ref{FQPIcons} and Fig.~\ref{F2} in the main text).  It is used only in panel B-C of Fig.~S\ref{FbulkQPI}-\ref{FXQPI}. 
 
The symmetry of QPI patterns is not only determined by the symmetry of the underlying electronic band structure, but also by the symmetry of the scattering potential responsible for generating the modulated electronic density. The $\mathbf{q}$-space QPI signal is a product of a bare QPI component multiplied by a structure factor component which, in the Born approximation, is the Fourier transform of the scattering potential.  The bare QPI is what would be generated by a rotationally symmetric point-scatterer \cite{Capriotti2003}. Experimentally, influence of the impurity structure factor has been observed by STM in Fe-based superconductors \cite{Allan2013a}: highly anisotropic QPI patterns were shown to be a direct consequence of anisotropic impurity states. 
 


\subsection{VI. Modelling the spectral function of a Kondo insulator}\label{specfunc}
Independent of the extraction of surface state trajectories from raw QPI data near the Fermi level already described, we use higher-energy QPI data to understand the bulk KI states. The ground state electronic structure of a Kondo insulator can be found by diagonalizing the Hamiltonian describing hybridization between an itinerant $d$-band, $\epsilon_d(\mathbf{k})$, and two closely spaced $f$-bands, $\epsilon_{f_1}(\mathbf{k}), \epsilon_{f_2}(\mathbf{k})$:
\begin{equation}\label{eq:hamilt}
\mathcal{H}(\mathbf{k}) = 
\left[
\begin{array}{ccc}
\epsilon_d(\mathbf{k}) &  -V_1(\mathbf{k})   & 
-V_2(\mathbf{k})  \\
-V_1(\mathbf{k})  &  \epsilon_{f_1}(\mathbf{k})  &  0 \\ 
-V_2(\mathbf{k})  &  0  &  \epsilon_{f_2}(\mathbf{k})
\end{array}
\right]
\end{equation}
where $V_i(\mathbf{k})$ is the momentum hybridization matrix element between $\epsilon_d$ and $\epsilon_{f_i}$ states.  Exclusion of the third crystal field split $f$ band, $\Gamma_7$, in our model is also consistent with calculations showing its minimal impact on the hybridized bands \cite{Takimoto2011,Lu2013}. In TKI theories for $\smb$, the expression for the hybridization, $V_i(\mathbf{k}) = v_i\sin(\mathbf{k}a_0)$ with $a_0$ the lattice constant,  is antisymmetric because of restrictions imposed by the opposite parity of the $d$ and $f$ states. The associated Kondo insulator spectral function and Green's function, follow from the standard definitions
\begin{align}
    \label{eq:green}
    \tilde{A}(\mathbf{k},E) &= -\mbox{Im}\tilde{G}(\mathbf{k},E)/\pi
    \nonumber
    \\
    \tilde{G}(\mathbf{k},E) &= \left[(E+i\mathbf{\Gamma}) \mathbbm{1}-\mathcal{H}(\mathbf{k})\right]^{-1}, \quad \mathbf{\Gamma} = [\Gamma_d, \Gamma_{f_1}, \Gamma_{f_2}]
\end{align}
where $\Gamma_i^{-1}$ are the quasiparticle lifetimes for the $i^{th}$ band and $\mathbbm{1}$ is the $3\times3$ identity matrix. 



The complexity of the Kondo insulator model necessitates optimizing the numerical algorithm for efficient computation time. One essential optimization converts all multidimensional integrals in momentum space to single dimensional ones by implementing a coordinate transformation, $k_z \rightarrow k_z^\prime$, to convert the ellipsoid constant energy contours of the light band, $\epsilon_d(\mathbf{k}) = E_i$, into spheres. The resulting isotropic band structure is then expressed as a function of a single radial coordinate, $k_r$. In this 1D representation the conduction band, $\epsilon_d(\mathbf{k})$, is modelled by a parabola with a band minimum at -1.6eV and a Fermi wavevector $k_F=0.54 \pi/a_0$, in good agreement with ARPES cuts along $\Gamma$-$X$ in momentum space \cite{Jiang2013}. The 1D representation reduces the 2D tight-binding Sm $4f$ bands in the model to an isotropic form,
\begin{equation}
    \epsilon_{f_i}(k_r) = \chi_{f_i}\cos (k_r a_0) - \mu_{f_i}
\end{equation}
where $2\chi_{f_i}$ is the bandwidth and $\mu_{f_i}$ is the center of the band with respect to the chemical potential. In the limit that the $f$ states are flat the 2D to 1D transformation does not affect the angular distribution of spectral weight so that $\mathbf{k}$-space integrals are unchanged.  With some bandwidth, undercounting of $f$ states may occur near the edges of the Brillouin zone.  One further optimization to our numerical techniques involved trials of various momentum dependent hybridization functions.  To within the resolution of our data we determined that the secondary hybridization term between $\epsilon_d$ and $\epsilon_{f_2}$ states could be replaced with a constant, $V_2(\mathbf{k}) = v_2$.  This simplification significantly sped up the fitting routine but maintained the same features as the use of $V_2(\mathbf{k}) = v_2\sin(\mathbf{k}a_0)$.  Our final model hybridized band structure closely matches first principle methods \cite{Lu2013} and provides an accurate description of the observed quasiparticle scattering dispersion in the higher energy range, away from the Fermi level where the independent surface state fitting was already described in section III. (Fig.\ S\ref{Ftrain}A-B). 
  \\
  
\subsection{VII. Co-tunneling formulation of STM differential conduction}

An electron from the STM tip can tunnel into a Kondo lattice through either the $d$-conduction or $f$-electron states at position ${\bf r}$.  Quantum interference between these paths means that STM differential conductance spectra, $g(E)$, can no longer be simply treated as proportional to the density of states. In this case, the expression to describe the experimentally measured spectra must be modified to include relative coupling to each electronic state component in addition to their interference terms \cite{Figgins2010a}.  In $\smb$, on-site hybridization between opposite parity $d$ and $f$ orbitals is suppressed and dominated by non-local interactions.  This non-locality gives rise to the antisymmetric form of the momentum space hybridization function, $V(\mathbf{k})$, crucial for the topological structure.  Consequently, co-tunneling into $d$ and $f$ states must occur via hybridization with neighboring sites.  However, recent numerical models show that non-local hybridization and co-tunneling in a TKI generates an effective local co-tunneling process into the $d$ and $f$ states with a $\mathbf{k}$-symmetric hybridization \cite{Baruselli2016}, motivating the local co-tunneling formulation we employ (equation (\ref{eq:didv}) of the main text). The expanded equation reads
\begin{widetext}
    \begin{multline}\label{eq:KIdidv}
    \frac{dI(\mathbf{r},E=eV)}{dV} = -\frac{2\pi e}{\hbar}N_\text{tip}(t_d)^2\bigg[ 
            \overbrace{N_{dd}(\mathbf{r},E) 
            + \left(\frac{t_{f_1}}{t_d}\right)^2 N_{f_1f_1}(\mathbf{r},E) 
            + \left(\frac{t_{f_2}}{t_d}\right)^2 N_{f_2f_2}(\mathbf{r},E)} ^ {\equiv D(E)} 
            \\ 
            + 2\frac{t_{f_1}}{t_d} N_{df_1}(\mathbf{r},E) 
            + 2\frac{t_{f_2}}{t_d} N_{df_2}(\mathbf{r},E) 
            + 2\frac{t_{f_1}}{t_d}\frac{t_{f_2}}{t_d}N_{f_1f_2}(\mathbf{r},E)\bigg].
    \end{multline}
\end{widetext}
where $N_{ij} = -\text{Im}\bar{G}_{ij}(\mathbf{r}-\mathbf{r}_0=0,E)$, and $\bar{G}_{ij}$ is the same as the Green's function defined in Methods Section VI but with the hybridization terms altered to be $\mathbf{k}$-symmetric.  Specifically, we chose the forms $V_1(\mathbf{k}) = v_1 |\sin(\mathbf{k})|$ and $V_2(\mathbf{k}) = v_2$ for the fits presented in this work (see previous section).  At each temperature, the fit to $g(E)$ is decomposed into the components of the above expression: three terms that directly couple to the density of states for the $d, f_1$ and  $f_2$ levels, called $D(E)$; and three interference terms (Fig.\ S\ref{Fqint}A-B). Here $D(E)$ is a weighted sum of the density of states for the three bands, and consequently reflects the opening of a gap as temperature is lowered (Fig.\ S\ref{Fqint}C). 


Fits of differential conductance spectra to equation (\ref{eq:KIdidv}) yields a complete parametrization of the Green's function in equation (\ref{eq:green}), subject to the functional forms of the various bands  described in the last section. The most important of these parameters are plotted as a function of temperature in Fig.~S\ref{FKIfit}.  Significant evolution is seen in the broadening term, $\Gamma_f$, of the two $f$ bands.  With increasing temperature, the states broaden with contributions from both thermal smearing and reduction of coherent scattering between the $f$ states themselves.  

\subsection{VIII. Dirac point confirmation by scanning tunneling spectroscopy}
\begin{table*}
 \caption{\label{tbl:ss} Comparison of surface state properties.}
  \begin{ruledtabular}
\begin{tabular}[c]{@{}lcccc@{}}
\toprule 
\textbf{Technique} & STM (present work) & Theory \cite{Lu2013}  & Photoemission \cite{Jiang2013} & Quantum oscillation \cite{Li2014}
\\ \hline
$\hbar v_{\bar{X}}$ (meV$\cdot$\AA) & $7.6 \pm 0.3$  &  $16 \pm 2$  &  $240 \pm 20$ & $1900 \pm 300$
\\
$E_{D_{\bar{X}}}$ (meV)  &  $-5.4 \pm 0.1$  & $1 \pm 1$   &  $-65\pm 4$  &  $-57 \pm 9$
\\
$2(k_{F_{\bar{X}}}-\bar{X}) (\pi/a_0)$  &  $0.54 \pm 0.02$  &  $0.25 \pm 0.02$    &  $0.38 \pm 0.03$ &  $0.039\pm 0.003$
\\
\hline
$\hbar v_{\bar{\Gamma}}$ (meV$\cdot$\AA) & $90 \pm 9$  &  $50 \pm 2$  &  $220 \pm 20$ & $4300 \pm 100$
\\
$E_{D_{\bar{\Gamma}}}$ (meV)  &  $-9 \pm 2$  &  $-7\pm 1$  &  $-23\pm 3$  &  $-460 \pm 20$
\\
$2k_{F_{\bar{\Gamma}}} (\pi/a_0)$  &  $0.07 \pm 0.01$  &  $0.18 \pm 0.02$    & $0.12 \pm 0.03$ &  $0.142\pm 0.001$
\\
\bottomrule
\end{tabular}
 \end{ruledtabular}
\end{table*}
The STM differential conductance spectra are decomposed into a surface and bulk contribution by fitting to the KI model in equation (\ref{eq:KIdidv}): the fit describes the bulk, while the residual is attributed to surface conduction (Fig.~S\ref{FSScons}).  This V-shaped residual matches expectations for topological surface states accessed by QPI: it is only appreciable for energies within the KI gap, and it has a nodal energy close to the Dirac point extracted independently from QPI (compare Fig.~S\ref{FSScons} and Fig.\ \ref{F2} in the main text).  Moreover, it is consistent across multiple samples, indicating that the QPI is not significantly influenced by the addition of a small concentration of dopants (perhaps because native Sm vacancies already act as magnetic scatterers; see Fig.~S\ref{Fsusc}). 

In a TKI, the temperature-dependent coherence of the bulk hybridization gap directly affects the surface states, in addition to thermal broadening. We simulated the effect of thermal broadening in three steps. First, we convolved measured $g(E)$ spectra with the derivative of the Fermi-Dirac distribution evaluated at the target temperature (Fig.~S\ref{FSSdecay}A). Second, we fit the resultant curve to the KI model (equation (\ref{eq:KIdidv})).  Finally, we computed the residual of the fit and its integrated spectral weight (Fig.~S\ref{FSSdecay}B).  We found that thermal broadening explains only a small fraction of the decay in the residual surface conductance as temperature is raised: the primary effect is from the bulk system losing coherence, which causes the Kondo screening to become unwound and the topological state to disappear (Fig.~S\ref{FSSdecay}C-D). 

\subsection{IX. Comparing STM, photoemission, and quantum oscillations}

The extraction of $\smb$ surface state properties by angle resolved photoemission \cite{Xu2014,Jiang2013,Neupane2013} and quantum oscillations experiments \cite{Li2014} has been controversial because of a large quantitative discrepancy between these techniques, in addition to disagreement with theoretical predictions. First, quantum oscillation velocities are one order of magnitude larger than those observed by photoemission. Second, the extracted velocities in photoemission experiments are approximately an order of magnitude larger when compared to first principles calculations \cite{Lu2013}. Furthermore, photoemission infers that the Dirac nodal energies lie well below the energy interval of the Kondo insulator gap, an observation that appears inconsistent with the topological origins of the surface states. Table II lists representative values of the surface state Fermi velocities determined by three experimental techniques and compares them to theory.  In contrast to extractions from quantum oscillation and photoemission experiments, the velocities measured in this work by STM are within a factor of two of theoretical predictions.

\end{document}